\documentclass{cupconf}
\usepackage{graphicx,psfig} 

  \checkfont{eurm10}
  \iffontfound
    \IfFileExists{upmath.sty}
      {\typeout{^^JFound AMS Euler Roman fonts on the system,
                   using the 'upmath' package.^^J}%
       \usepackage{upmath}}
      {\typeout{^^JFound AMS Euler Roman fonts on the system, but you
                   dont seem to have the}%
       \typeout{'upmath' package installed. cupconf.cls can take advantage
                 of these fonts,^^Jif you use 'upmath' package.^^J}%
      }
  \else
  \fi

  \checkfont{msam10}
  \iffontfound
    \IfFileExists{amssymb.sty}
      {\typeout{^^JFound AMS Symbol fonts on the system, using the
                'amssymb' package.^^J}%
       \usepackage{amssymb}%
       \let\le=\leqslant  
       \let\ge=\geqslant  
      }{}
  \fi

  \IfFileExists{amsbsy.sty}
    {\typeout{^^JFound the 'amsbsy' package on the system, using it.^^J}%
     \usepackage{amsbsy}}
    {}

\newsavebox{\astrutbox}
\sbox{\astrutbox}{\rule[-5pt]{0pt}{20pt}}

\title[Parameters and Winds of Hot Massive Stars]{Parameters and Winds of
Hot Massive Stars}

\author[R. P. Kudritzki \& M.A. Urbaneja]%
{R\ls O\ls L\ls F\ns P.\ns K\ls U\ls D\ls R\ls I\ls T\ls Z\ls K\ls I\break
\and M\ls I\ls G\ls U\ls E\ls L\ns A.\ns U\ls R\ls B\ls A\ls N\ls E\ls J\ls A}

\affiliation{Institute for Astronomy, University of Hawaii,
2680 Woodlawn Drive, Honolulu, HI 96822, USA}

\pubyear{2006}
\volume{STScI May Symp. 2006}
\pagerange{119--126}
\date{July 15, 2006}
\begin{document}

\maketitle

\begin{abstract}

Over the last years a new generation of model atmosphere codes, which
include the effects of metal line-blanketing of millions of spectral
lines in NLTE, has been used to re-determine the properties of massive
stars through quantitative spectral analysis methods applied to optical,
IR and UV spectra. This has resulted in
a significant change of the effective temperature scale of early type stars 
and a revision of mass-loss rates. Observed mass-loss rates and effective
temperatures depend strongly on metallicity, both in agreement with
theoretical predictions. The new model atmospheres in conjunction with the
new generation of 10m-class telescopes equipped with efficient multi-object
spectrographs have made it possible to study blue supergiants in galaxies
far beyond the Local Group in spectroscopic detail to determine accurate
chemical composition, extinction and distances. A new distance determination
method, the flux weighted gravity - luminosity relationship, is discussed as
a very promising complement to existing stellar distance indicators.

Observationally, there are still fundamental uncertainties in the
determination of stellar mass-loss rates, which are caused by the fact
that there is evidence that the winds are inhomogeneous and clumped. This
may lead to major revisions of the observed rates of mass-loss.

\end{abstract}

\firstsection 
\section{Introduction}

Hot massive stars are cosmic
engines of fundamental importance not only in the local but also in
the early universe. A first generation of very massive stars has very likely
influenced the formation and evolution of the first building blocks
of galaxies. The spectral appearance of Lyman break galaxies and 
Ly$\alpha$-emitters at high redshift is dominated by an intrinsic
population of hot massive stars. Gamma-ray bursters are very likely the 
result of terminal collapses of very massive stars and may allow to trace
the star formation history of the universe to extreme redshifts.

It is obvious that the understanding of important processes of star and
galaxy formation in the early universe is intimately linked to our 
understanding of the physics of massive stars. The observational constraints
of the latter are provided by quantitative spectroscopic diagnostics of the
population of hot massive stars in the local universe. It is the goal of
this contribution to provide an overview about the dramatic progress, which
has been made in this field over the last five years.

There are two factors which have contributed to this progress, new
observational facilities such as the optical/infrared telescopes of the
10m-class on the ground and observatories in space allowing for spectroscopy
in the UV (HST, FUSE, GALLEX), IR (ISO, Spitzer) and at X-ray wavelengths (XMM,
Chandra) and the enormous advancement of model atmosphere and radiative
transfer techniques. As for the latter, it is important to realize that
modelling the atmospheres of hot stars is a tremendous challenge. Their
physics are complex and very different from standard stellar atmosphere
models. They are dominated by a radiation field with energy densities
larger than or of the same order as the energy density of the atmospheric
matter. This has two important consequences. First, severe departures from 
Local Thermodynamic Equilibrium of the level populations in the entire 
atmosphere are induced, because radiative transitions between ionic energy 
levels become much more important than inelastic collisions. Second, 
supersonic hydrodynamic outflow of atmospheric matter is initiated by line 
absorption of photons transferring outwardly directed momentum to the 
atmospheric plasma. This latter effect is responsible for the existence of 
the strong stellar winds observed and requires the use of NLTE model
atmospheres, which include the hydrodynamic effects of stellar winds (see
\cite{kud98} for a detailed description of the physics of hot star
atmospheres).

The winds of hot massive stars are fundamentally important. Their energy and
momentum input into the ISM is significant creating circumstellar shells, 
stellar wind bubbles and initiating further star formation. They affect 
the stellar evolution by modifying evolutionary timescales, chemical
profiles, surface abundances and stellar luminosities. They also have
substantial effects on the structure of the stellar atmospheres. They dominate
the density stratification and the radiative transfer through their
transonic velocity fields and they modify the amount of the
emergent ionizing radiation significantly (Gabler et al., 1989, 1991, 1992,
Najarro et al., 1996).

\begin{figure}
\centerline{\hbox{
   \psfig{figure=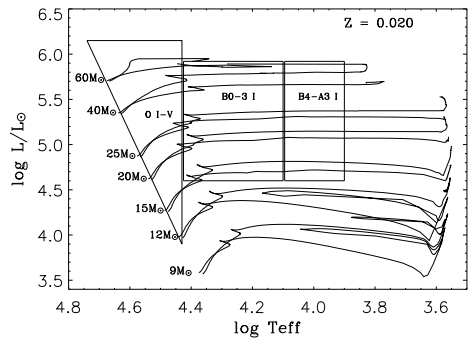,width=7cm}
     }}
\caption[]{
Evolutionary tracks of massive stars in the HRD with (solid) and without
(dashed) stellar rotation, \cite{meynet00}) and the domains of 
spectral types discussed in this review.
\label{hrd}}	 
\end{figure}

While the basic concepts for the hydrodynamic atmospheres of hot stars and
the spectroscopic diagnostics of their parameters and stellar winds have
been developed in the eighties and nineties (see reviews by \cite{kuhu90},
\cite{kud98} and \cite{kupu00}), the development of the most recent generation of
model atmospheres, which for the first time accounts selfconsistently for the
effects of NLTE metal line-blanketing (see section 2), has lead to a dramatic
change of the diagnostic results. A new effective temperature scale has been
obtained for the O-star spectral types, which are apparently significantly
cooler than originally assumed and, thus, on average have a lower luminosity
and mass and also provide less ionizing photons, than previously thought. We
will discuss these effects in detail in section 2, 3, and 4. Note that we will
focus our review on ``normal'' hot massive stars in a mass-range between
15 and 100 $M_{\odot}$ in well-established evolutionary stages such as
dwarfs, giants, and supergiants (Fig.\,\ref{hrd}). We will not discuss 
objects with extreme
winds such as Wolf-Rayet stars, luminous blue variables in outbursts, etc.
For those, we refer the reader to the contributions by Paul Crowther and
Nathan Smith in this volume.

The new model atmospheres in conjunction with the
new generation of 10m-class telescopes equipped with efficient multi-object
spectrographs allow to study blue supergiants in galaxies far beyond the 
Local Group in great spectroscopic detail to determine accurate
chemical composition, extinction and distances. We will report recent
results in this new field of ``extragalactic stellar astronomy'' in section 5.
In section 6, we will discuss a new distance determination method based on
stellar photospheric spectroscopy, the flux weighted gravity - luminosity 
relationship.

The diagnostics of stellar winds, in particular of mass-loss rates, are also
affected by the proper accounting for line-blanketing affects. While the
general scaling relations of stellar wind parameters with stellar
luminosity, mass, radius, and metallicity as described in \cite{kupu00} 
remain qualitatively unchanged, important quantitative changes have been found.
There are still puzzling uncertainties with regard to the observed
rates of mass-loss related to the inhomogeneous structure of the stellar wind
outflows. We will present and discuss the most recent results in section 7.

\begin{figure}
\centerline{\hbox{
   \psfig{figure=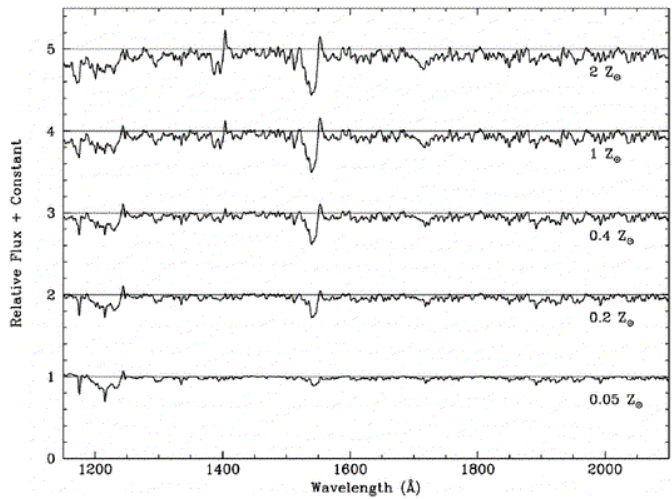,width=9cm}
     }}
\caption[]{
Model UV spectra of an integrated population of massive stars in a starburst
galaxy displayed as a function of metallicity. The model atmosphere code used
was WMBasic (\cite{pauldrach01}) and the population synthesis was done with
Starburst99. (From \cite{rix04}).
\label{starburst}}	 
\end{figure}

As indicated above, the new generation of model atmosphere codes has
important applications on the interpretation of spectra of
star-bursting galaxies in the early universe. Fig.\,\ref{starburst} is a
nice example. This work by \cite{rix04} has been used to constrain star
formation rates and metallicities in high redshift Lyman-break galaxies.
Another example is the work by \cite{barton04}, which uses stellar
atmosphere model predictions by \cite{bromm01} and \cite{schaerer03} to
explore the possibility to detect the first generation of very massive stars at
redshifts around ten with present day 8m-class telescopes and with future 
(diffraction limited) optical/IR telescopes of 30m aperture such as the 
GSMT (see also report by the GSMT Science Working Group, \cite{kud03a}).
As it turns out, not only would the GSMT be able to detect such objects it
would also allow to constrain the IMF of massive stars in the early universe
from the relative comparison of L$_{\alpha}$ and HeII1640 recombination
lines caused by the population of the first stars.

\section{The effects of metal line blanketing}

\begin{figure}
\centerline{\hbox{
   \psfig{figure=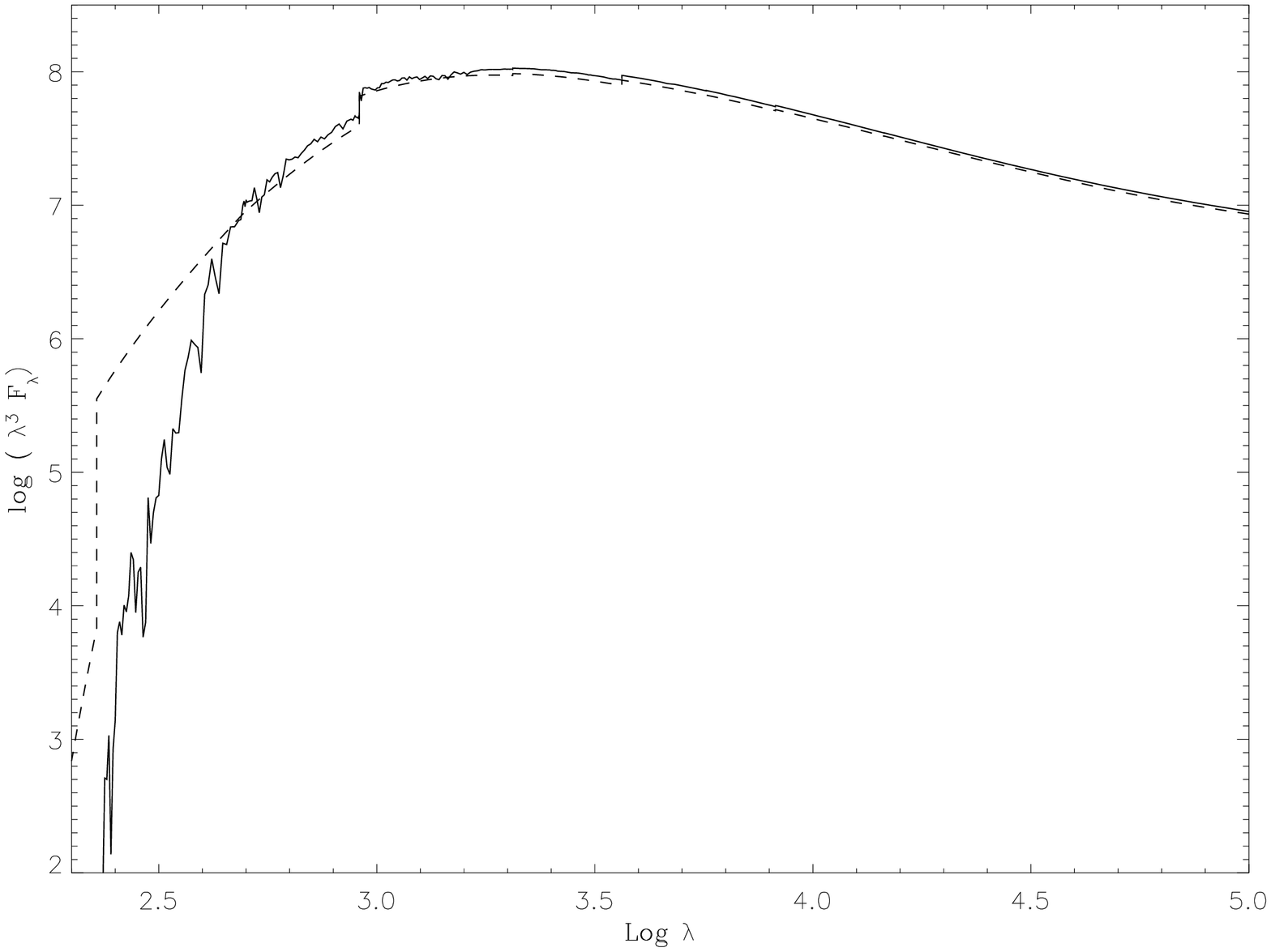,width=7cm,angle=90}
   \psfig{figure=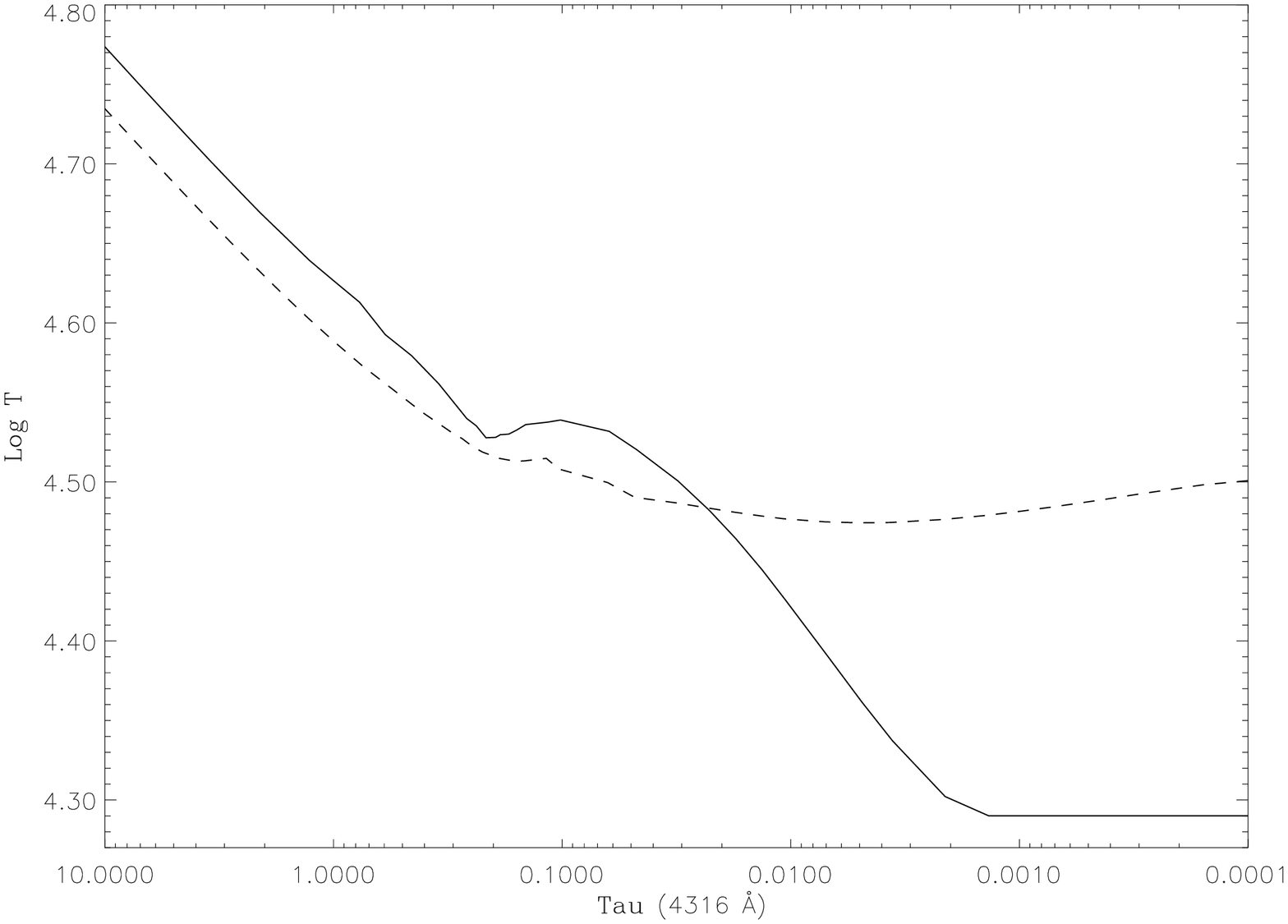,width=7cm,angle=90}
         }}
\caption[]{
Effects of metal line blanketing. Left: Emergent flux of a NLTE
hydrodynamic O-star model of T$_{eff}$=40,000K with (solid) and without
(dashed) metal line blanketing. Right: Local kinetic temperature as a
function of monochromatic optical depth at blue wavelength for the same two
models. The photospheric backwarming as discussed in the text as the result
of blanketing is clearly seen. The NLTE code FASTWIND (\cite{puls05}) was 
used for the calculations.
\label{flux}}
\end{figure}

\begin{figure}
\centerline{\hbox{
   \psfig{figure=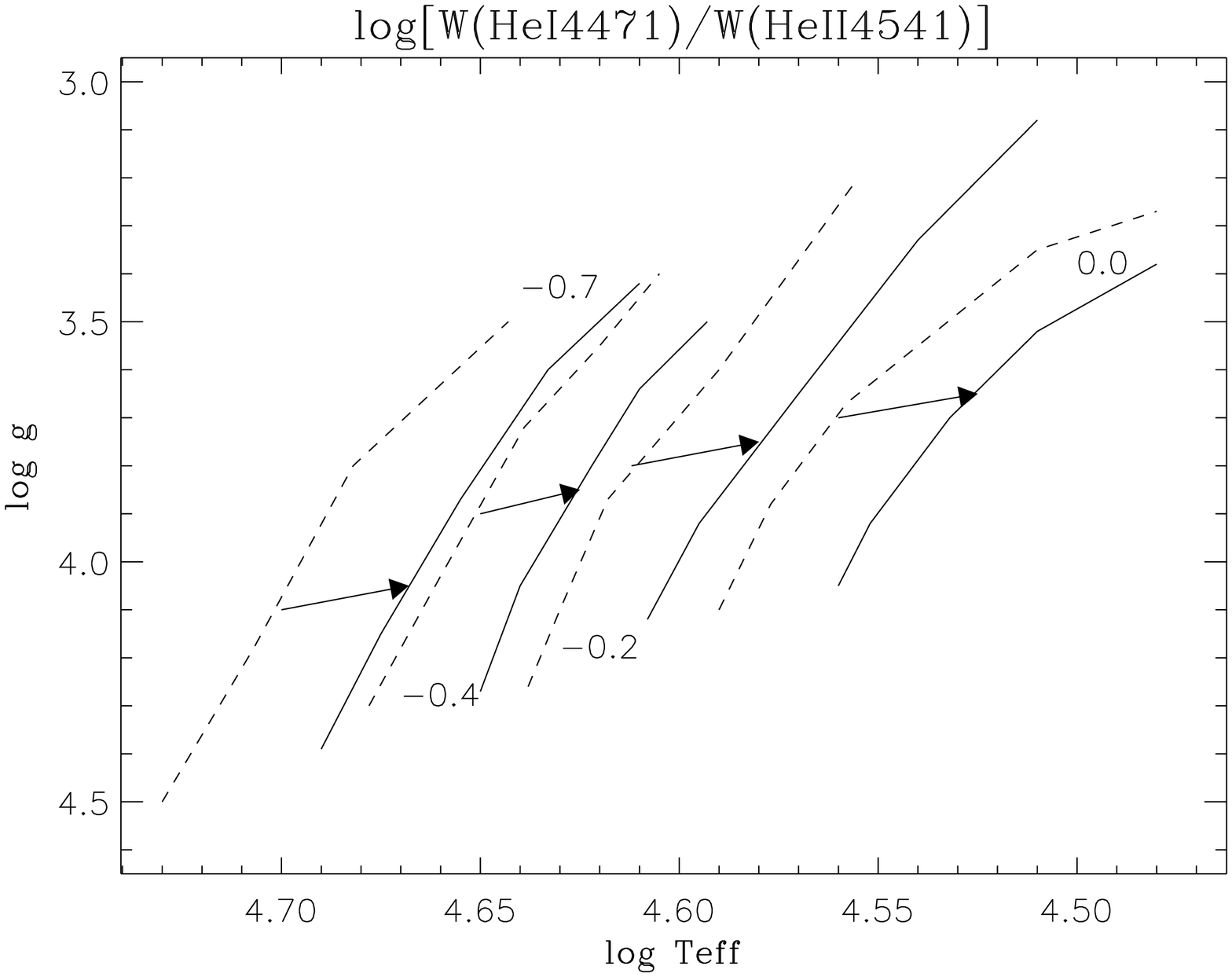,width=7cm,angle=90}
   \psfig{figure=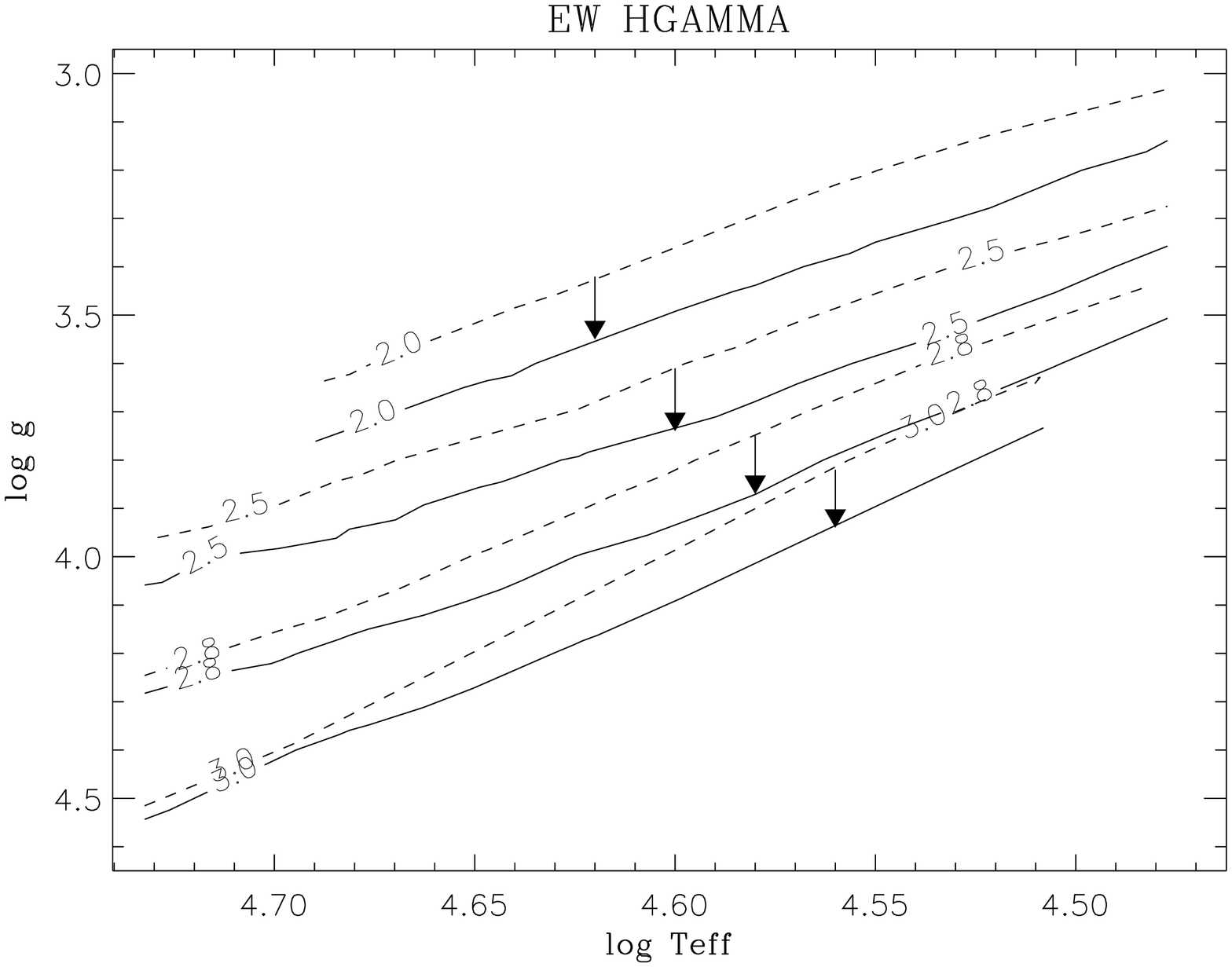,width=7cm,angle=90}
         }}
\caption[]{Effects of metal line blanketing on the stellar diagnostics.
Plotted are the isocontours of  
log [W$_{\lambda}$(HeI4471)/W$_{\lambda}$(HeII4542)] (left) and
W$_{\lambda}$(H$_{\gamma}$) (right) in the (log g, log T$_{eff}$) - plane.
Dashed isocontours are unblanketed models, solid are blanketed. Vectors
indicate the shifts caused by the effects of NLTE metal line blanketing. 
As for the previous figures the calculations were done with the NLTE code 
FASTWIND (\cite{puls05}).
\label{diag}}
\end{figure}

The inclusion of the opacity of millions of spectral lines in NLTE has two
major effects. First, it changes the spectral energy distribution in
the UV because of strong metal line absorption in the outer atmosphere
(``line-blanketing''). An example is given in Fig.~\ref{flux}. While the UV
flux is reduced, the physical requirement of radiative equilibrium and
conservation of the total flux leads to an increase of the emergent flux at
optical and IR wavelengths relative to the unblanketed case. This increase
comes from the fact that a significant fraction of the photons absorbed in 
the outer layers are 
emitted back to the inner photosphere providing additional energy input 
and, thus, heating the deeper photospheres. This second ``backwarming''
effect increases the local kinetic temperature (see Fig.~\ref{flux}),
which then leads to stronger photospheric continuum emission and
modifies diagnostically important ionization equilibria such as HeI/II, 
which are used for the
determination of T$_{eff}$. Fig.~\ref{diag} demonstrates how the HeI/II
ionization equilibrium is shifted towards lower T$_{eff}$ because of the
backwarming effect. At the same time, the pressure-broadened wings of the
Balmer lines (the standard diagnostic for log g) become weaker, because
the millions of metal lines increase the radiative acceleration g$_{rad}$
and decrease the effective gravity g$_{eff}$ = g - g$_{rad}$. As a result
higher gravities are needed to fit the Balmer lines (see Fig.~\ref{diag})
in addition to the lower temperatures obtained from the helium ionization
equilibrium. In summary, the use of blanketed models leads to systematic
shifts in the (log g, log T$_{eff}$) - plane towards lower temperatures and
higher gravities. For stars with known distances to be analyzed this means
lower luminosities and lower masses. We also note that the presence
of dense stellar wind envelopes increases the effects of backwarming 
(see \cite{hummer82}, \cite{abbott85}). In addition, stellar winds also 
affect the strengths of diagnostically crucial HeI absorption lines by 
contaminating the absorption with stellar wind line emission (\cite{gab89},
\cite{sellmaier93}, \cite{repo04}).

The combined effects of line blanketing and backwarming affect also the
ionizing fluxes. Amazingly, for the ionization of hydrogen the changes are
very small as the effects of blanketing and backwarming balance each other.
However, the ionization of ions with absorption edges shorter than the one
for hydrogen is significantly affected (see \cite{kud02}, \cite{martins05a}).

Over the last years very powerful and user friendly software packages 
to calculate NLTE model atmospheres with metal line blanketing have been
developed, which are now available to the astronomical
community and which have been intensively applied for spectral diagnostics
of hot massive stars. At Munich University
Observatory the codes ``WMBasic'' (\cite{pauldrach01}) and ``FASTWIND''
(\cite{puls05}) were developed in partial collaboration with the authors of
this article and at Pittsburgh the code ``CMFGEN'' is the result of
intensive work by John Hillier and collaborators (\cite{hillier03}). Ivan
Hubeny and Thierry Lanz have developed ``TLUSTY'' (\cite{lanz03}), which
does not include the hydrodynamics of winds but can be applied in all cases,
where the stellar winds are weak.

\begin{figure}
\centerline{\hbox{
   \psfig{figure=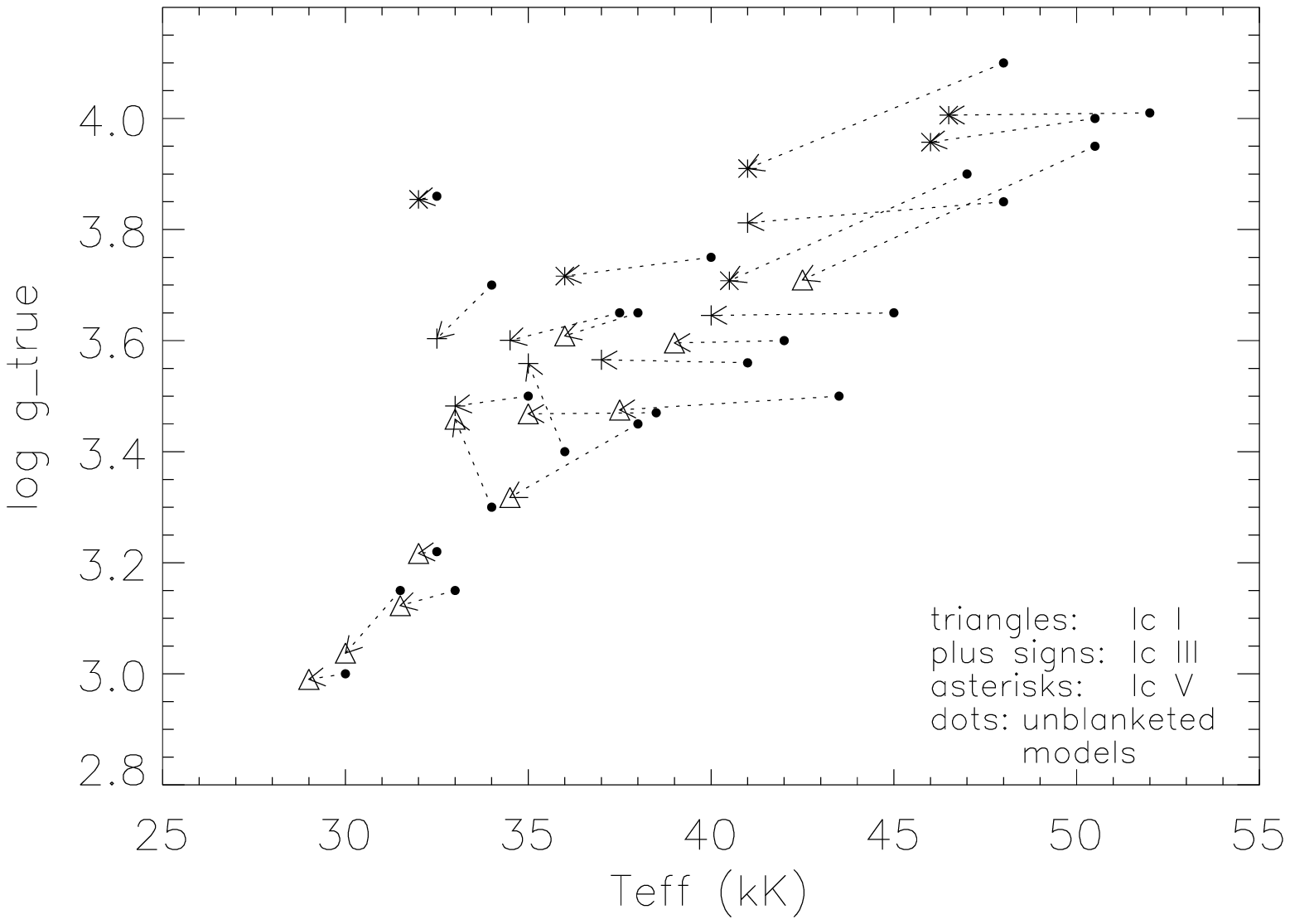,width=7cm}
   \psfig{figure=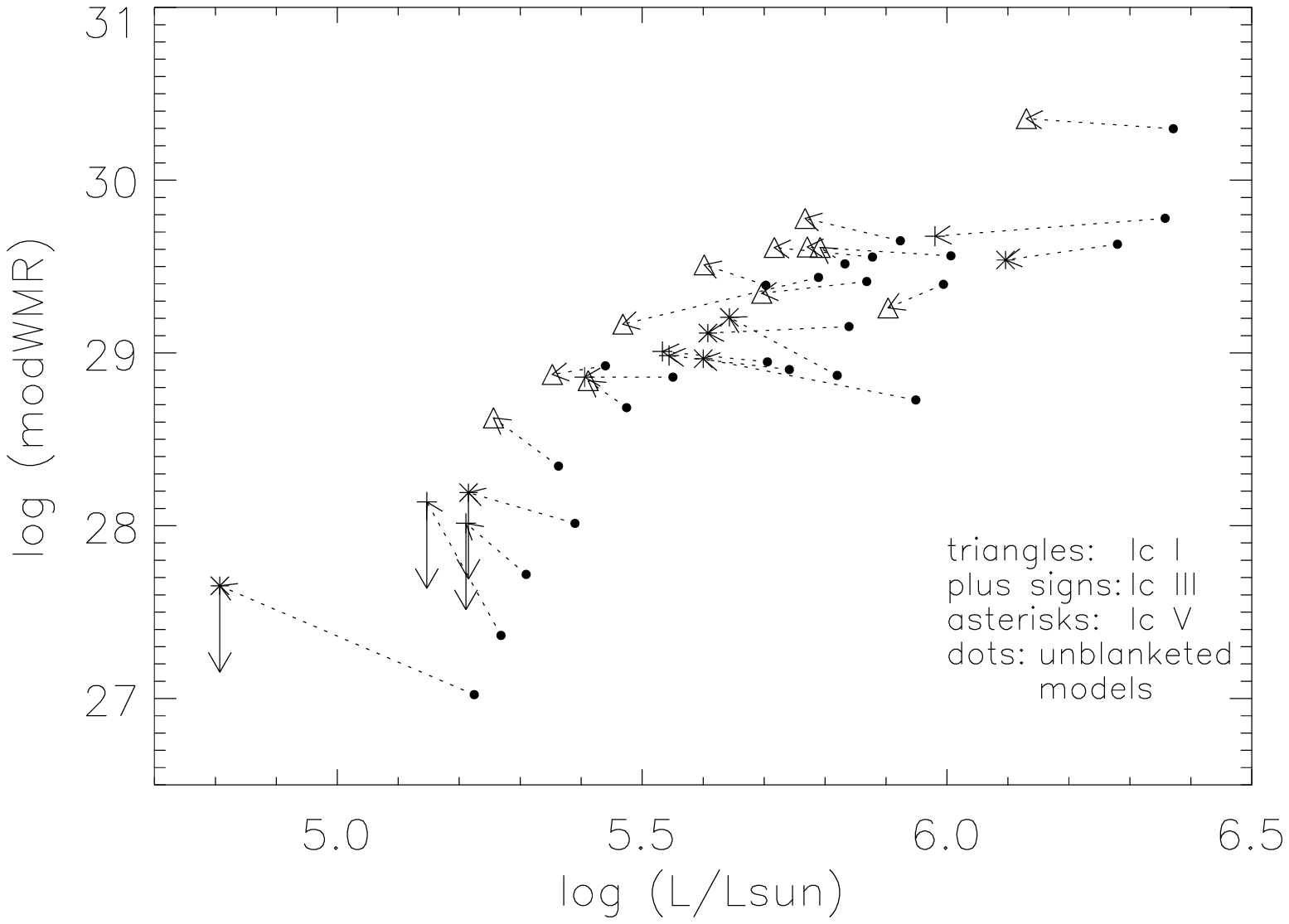,width=7cm}
         }}
\caption[]{Effects of metal line blanketing on the stellar diagnostics.
Left: Shifts of the location of individual O-stars in the 
(log g, log T$_{eff}$)-plane caused by the use of metal line blanketed models 
in the analysis of the hydrogen and helium optical line spectrum.
Right: Shifts of modified stellar wind momenta (see section 7)
as a function of luminosity. (From \cite{repo04}).
\label{odiag1}}
\end{figure}

\section{The effective temperature scale of O-stars}

With the new generation of NLTE metal line blanketed model atmospheres 
available quite a number of very detailed spectroscopic studies of O-stars
have been published over the last five years. While the diagnostic methods
used are generally still the same as in the earlier work by \cite{kud80},
\cite{kud83}, \cite{kud89}, \cite{kuhu90}, \cite{herrero92}, which
was based on a first generation of hydrostatic NLTE model atmospheres with
hydrogen and helium opacity only, the results coming out of the new work are
substantially different because of the tremendous improvements of the model
atmospheres. Since the effects of metal line blanketing depend on the
stellar metallicity, we will discuss O-stars in the solar neighborhood of
the Milky Way and in the Magellanic Clouds separately.

\subsection{O-stars in the Milky Way}

Milky Way O-stars have been studied by \cite{pauldrach01},
\cite{herrero02}, \cite{martins02}, \cite{bianchi02},
\cite{garcia04}, \cite{repo04}, \cite{markova04}, \cite{martins05a},
\cite{mokiem05}, \cite{bouret05}. The result of all these studies with
regard to the effective temperature scale is quite dramatic. An example is
given by Fig.~\ref{odiag1}, which displays the shifts in the (log g, log
T$_{eff}$)-plane of indivual O-stars, when the classical analysis of
hydrogen and helium lines based on unblanketed model atmospheres is replaced
by blanketed models. As to be expected from the previous section, the use of
blanketed models leads to cooler effective temperatures, which is simply
caused by the fact that blanketed models have higher intrinsic local 
photospheric temperatures and thus models with lower T$_{eff}$ are required
to fit the observed helium ionization equilibrium. In addition, blanketed
models have a lower local gas pressure and, thus, a higher gravity is generally
needed to fit the pressure broadened Balmer lines. 

\begin{figure}
\centerline{\hbox{
   \psfig{figure=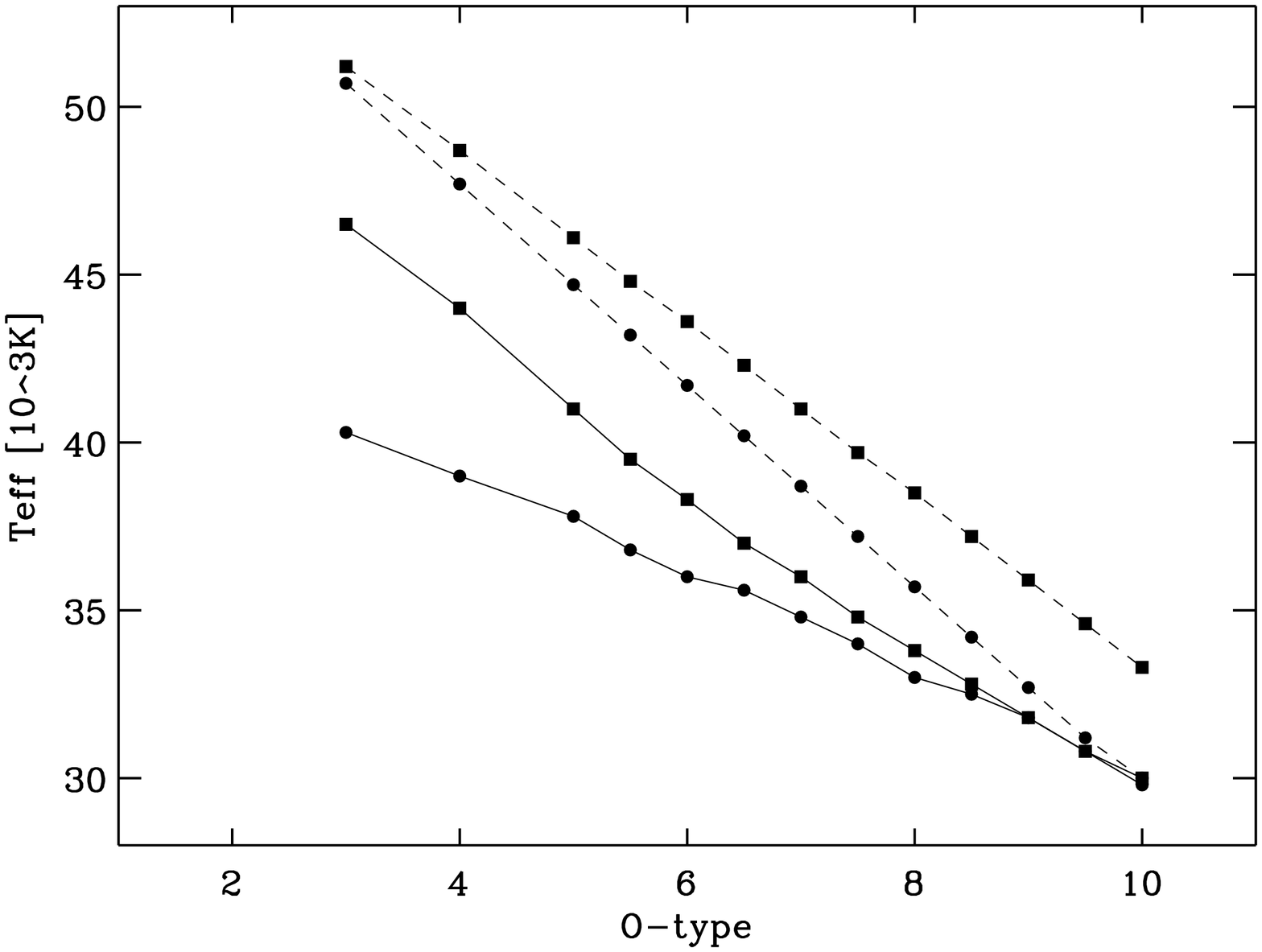,width=7cm,angle=90}
   \psfig{figure=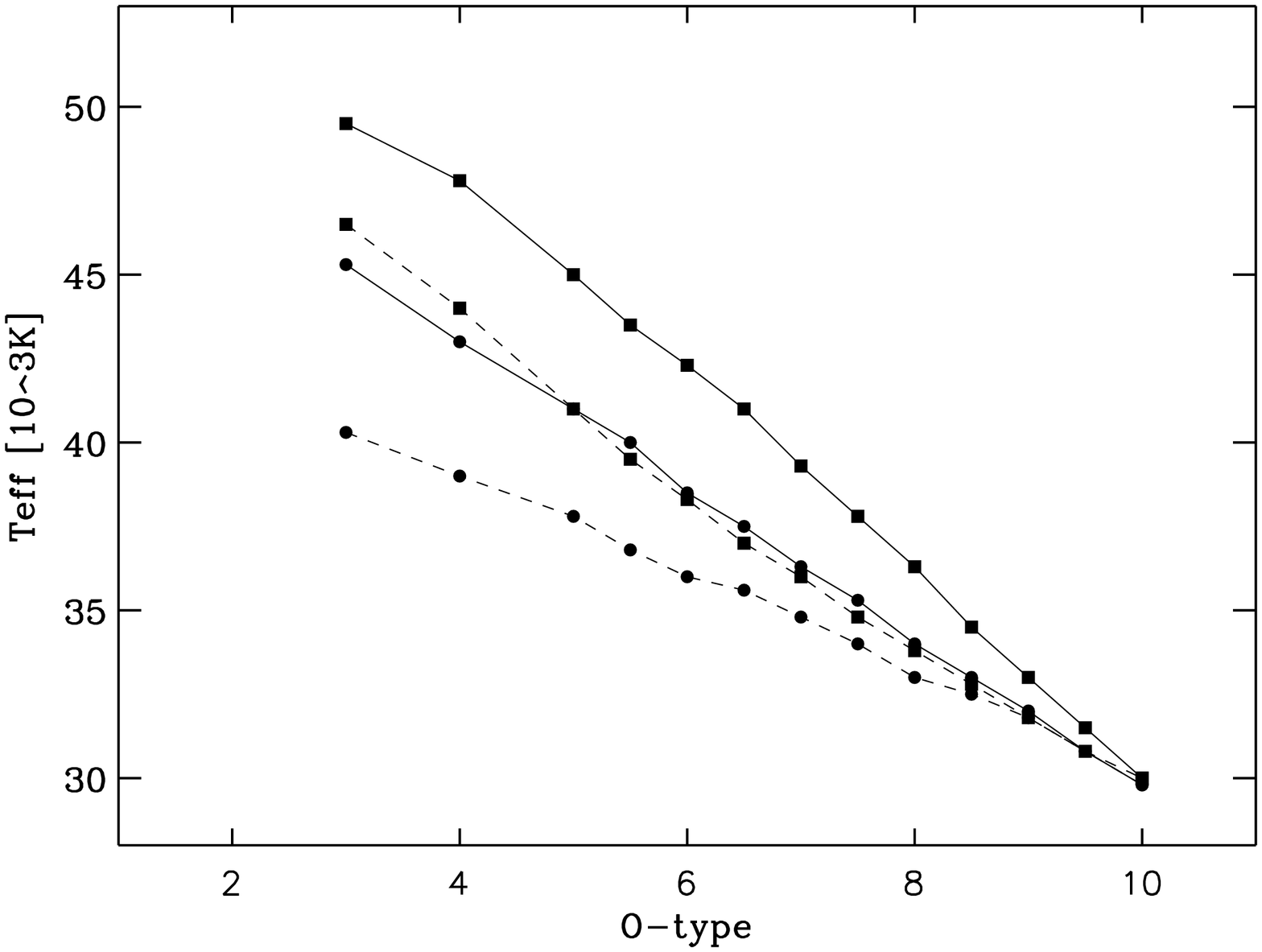,width=7cm,angle=90}
         }}
\caption[]{Left: Effective temperature as a function of spectral type for Milky
Way O-stars. Supergiants are plotted as circles and giants and dwarfs as
squares. The new scales are represented by the solid curves,
and the old scales by the dashed curves. For discussion see text.
Right: The new effective temperature scale of O-stars in the SMC (solid)
relative to the Milky Way (dashed). (From \cite{massey05}).
\label{mwomass}}
\end{figure}

While systematic effects towards somewhat lower effective temperatures and
higher gravities, were always expected in previous work based on unblanketed
models, the large shifts as displayed in Fig.~\ref{odiag1} come as a surprise.
We note that the effects are strongest for very hot objects and low gravity
supergiants, which have very strong and dense winds affecting the helium
ionization equilibrium additionally by the mechanisms discussed above. In
some cases we encounter effective temperature changes of the order of 10 to
15 percent. Based on the work by \cite{repo04}, \cite{massey05}
have introduced a new effective temperature scale for Milky
Way O-stars, which is displayed in Fig.~\ref{mwomass} and compared to the
old scale by \cite{vacca96}. We realize that, in particular for supergiants,
the differences between the two scales are dramatic. A very similar new
effective temperature scale has also been introduced independently by 
\cite{martins05a} based on a comprehensive study of a large sample of O-stars.

\subsection{Quantitative IR spectroscopy of O-stars}

\begin{figure}
\centerline{\hbox{
   \psfig{figure=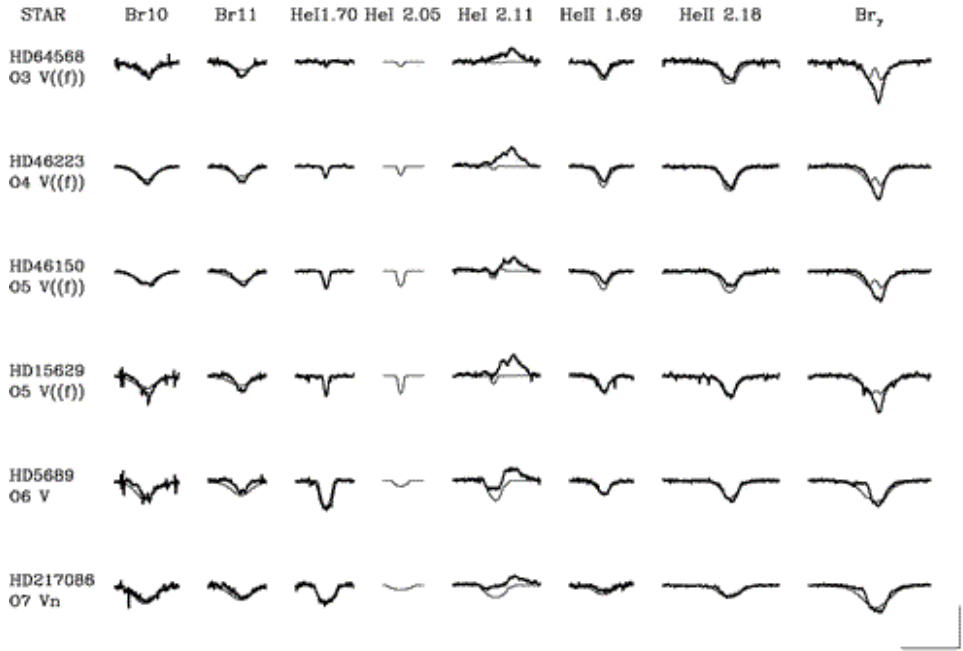,width=13cm}
         }}
\caption[]{
Model atmosphere fits of IR hydrogen and helium lines to determine stellar
parameters. Note that HeI 2.11$\mu$ is blended by nitrogen lines not
included in the calculations. Br$_{\gamma}$ is blended by HII region
emission in some cases. (From \cite{repo05}).
\label{ir}}
\end{figure}

Massive stars are frequently found in star forming regions heavily obscured
by interstellar dust. In such a situation IR spectroscopy is the only way to
obtain information. In the galactic center \cite{naj94}, \cite{naj97},
\cite{figer98}, \cite{naj04} (see also the contribution by Don Figer in this 
volume) have demonstrated how quantitative IR spectroscopy can be used to 
determine the stellar parameters of very extreme supergiants
and Wolf-Rayet stars. Similar work has now been carried more recently for
``normal'' O-stars with the goal to find out
whether the analysis of the hydrogen and helium lines in the IR yields
accurate information about stellar parameters consistent with the ones
obtained from the analysis of the optical spectrum. \cite{hanson05} and
\cite{repo05} have used Subaru H and K band spectra of high S/N to study a
large sample of O-stars with very encouraging results showing that IR
spectroscopy leads to effective temperatures, gravities and mass-loss rates
practically identical with those determined from optical spectra
(Fig.~\ref{ir} gives an example). In parallel, \cite{lenorzer04} have 
carried out a systematic model atmosphere study of the IR spectral 
diagnostics for normal O-stars focussing on both stellar lines as well as
HII region emission lines. It is obvious from this new work that 
for future investigations of massive stars
in dense and highly obscured star forming regions quantitative IR
spectroscopy has an enormous potential.

\subsection{Effects of metallicity. The effective temperature scale of
O-stars in the Magellanic Clouds}

\begin{figure}
\centerline{\hbox{
   \psfig{figure=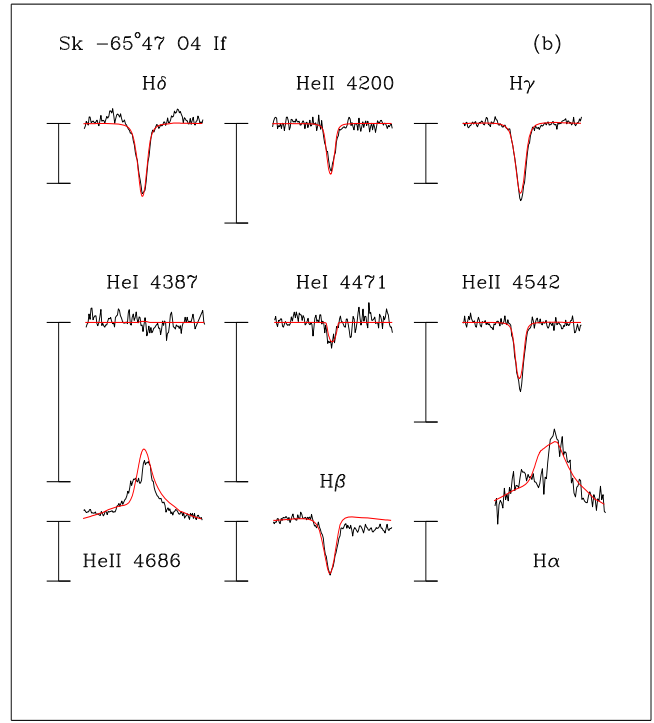,width=10cm}
         }}
\caption[]{
Model atmosphere fits of optical hydrogen and helium lines of an O-star 
in the LMC to determine stellar parameters. (From \cite{massey04}).
\label{mcospec}}
\end{figure}

From the discussion in section 2 it is clear that metallicity must have an
influence on the strengths of blanketing effects and, thus, the effective
temperature scale of O-stars. The ideal laboratory to investigate the 
metallicity dependence are the Magellanic Clouds because of their lower
metallicity relative to the Milky Way. \cite{massey04} and \cite{massey05}
and, independently, \cite{mokiem04} and \cite{mokiem06a} have carried out 
a systematic and comprehensive spectroscopic study of O-stars in the clouds.
Fig.~\ref{mwomass} and Fig.~\ref{mcospec} describe the work done by Massey
and collaborators. From a detailed fit of the optical hydrogen and helium
lines and HST and FUSE UV spectra they were able to determine stellar
parameters and stellar wind properties. The result with regard to the
effective temperature scale is striking and in agreement with the
expectation based on model atmosphere theory. For a given spectral type
defined by the relative strengths of HeI and HeII lines O-stars at lower
metallicity are significantly hotter than their galactic counterparts, a
result which is also clearly confirmed by the work of Mokiem and collaborators.
For the determination of IMFs in other galaxies based on spectral
classification or
for the investigation of the stellar content of galaxies or HII-regions
based on nebular emission line analysis this is an important effect to be
taken into account. Also for the study of integrated spectra of starburst
galaxies the metallicity dependence of the effective temperature scale might
be important.

\subsection{Systematic effects depending on the analysis method}

\begin{figure}
\centerline{\hbox{
   \psfig{figure=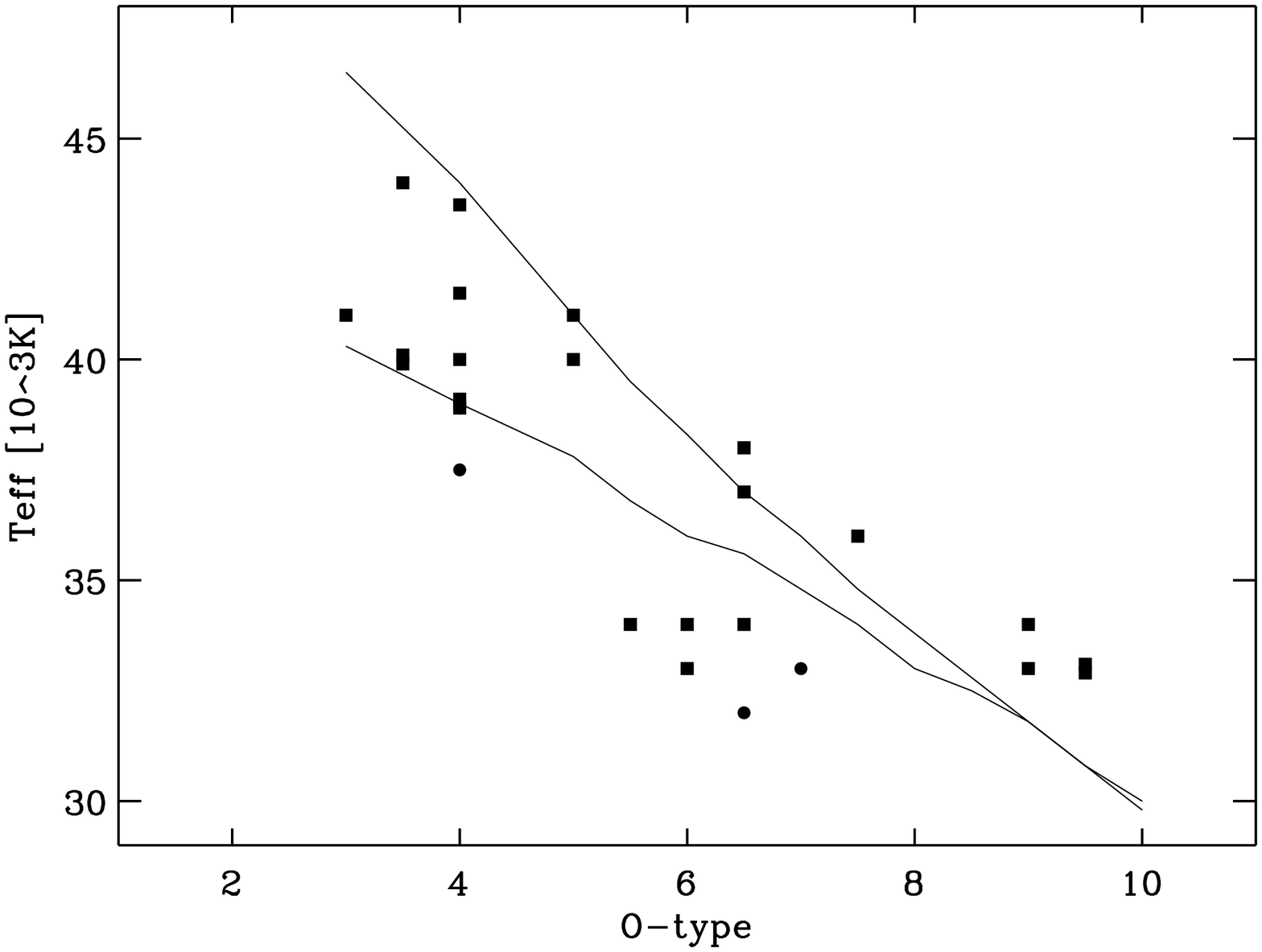,width=7cm,angle=90}
   \psfig{figure=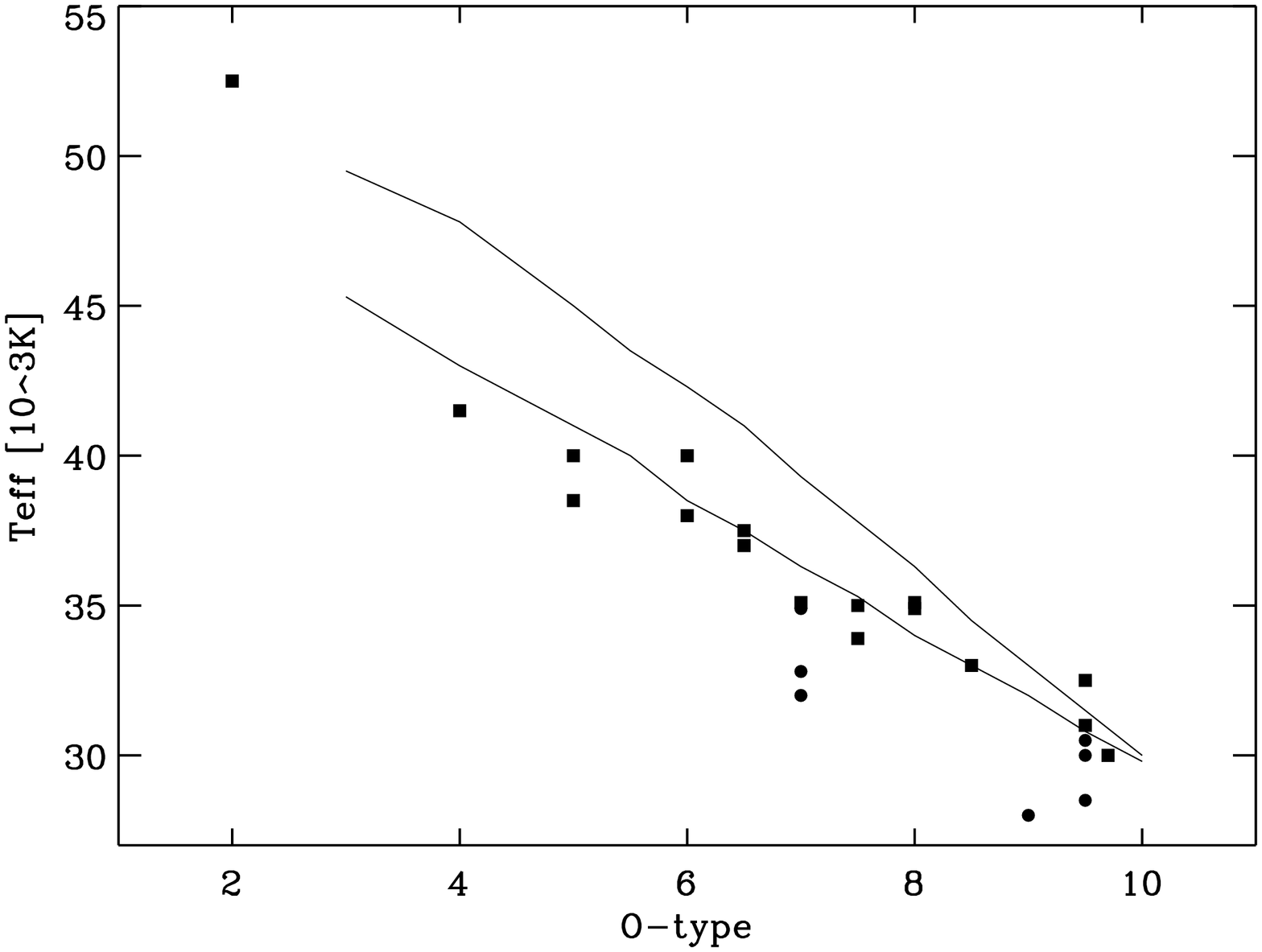,width=7cm,angle=90}
         }}
\caption[]{
O-star effective temperature scales according to \cite{massey05} for the
Milky Way (left) and the the SMC (right) based on the analysis of the HeI/II
ionization equilibrium at optical wavelengths. Overplotted are the results
of spectral analyses using UV metal lines as temperature indicator (see text
for the data sources). Supergiants are plotted as circles giants and 
dwarfs as squares.
\label{uvtemp}}
\end{figure}

In the work discussed so far effective temperatures and gravities have been
determined from the fit of the He I/II ionization equilibrium of optical or
IR helium lines and the Balmer lines. Since O-stars are UV bright and
hundreds of photospheric lines can be identified in high resolution UV
spectra (IUE, HST, ORFEUS, FUSE), it is tempting to use this spectral range
to obtain independent information about the stellar parameters. As it turns
out, it is difficult to constrain gravities solely through the UV, however,
there are ionization equlibria such as Fe IV/V/VI or CIII/IV which can be
used for a determination of temperatures. For Milky 
Way O-stars such work has been carried out by \cite{pauldrach01}, 
\cite{bianchi02}, \cite{garcia04}, and \cite{bouret05}. As shown in
Fig.~\ref{uvtemp} significantly lower T$_{eff}$ are obtained in some cases
from the UV lines compared to optical/IR HeI/II lines. In particular, the
work by \cite{bianchi02} and \cite{garcia04} has resulted in very low
effective temperatures.

A similar result has been obtained by \cite{heap06} for SMC O-stars mostly
of luminosity class V. Fig.~\ref{uvtemp} shows their results compared to
the effective temperature scale obtained by \cite{massey04} and
\cite{massey05}. Fig.~\ref{uvtemp} also displays the results found by
\cite{crowther02}, \cite{hillier03}, \cite{bouret03}, and \cite{evans04},
which show a less extreme but similar trend. 

This indication of a systematic effect depending on the analysis method
deserves a very careful and systematic future investigation. What is needed 
is a
comprehensive simultaneous UV, optical, IR analysis of all the O-stars
studied so far. The observational material seems to be available
or can be easily obtained. Of course, this will be a time consuming effort
requiring a lot of detailed spectroscopic and model atmosphere work.
However, it is extremely important that the reasons for these obvious
discrepancies are well understood.

\section{The effective temperature scale of B and A supergiants}

\begin{figure}
\centerline{\hbox{
   \psfig{figure=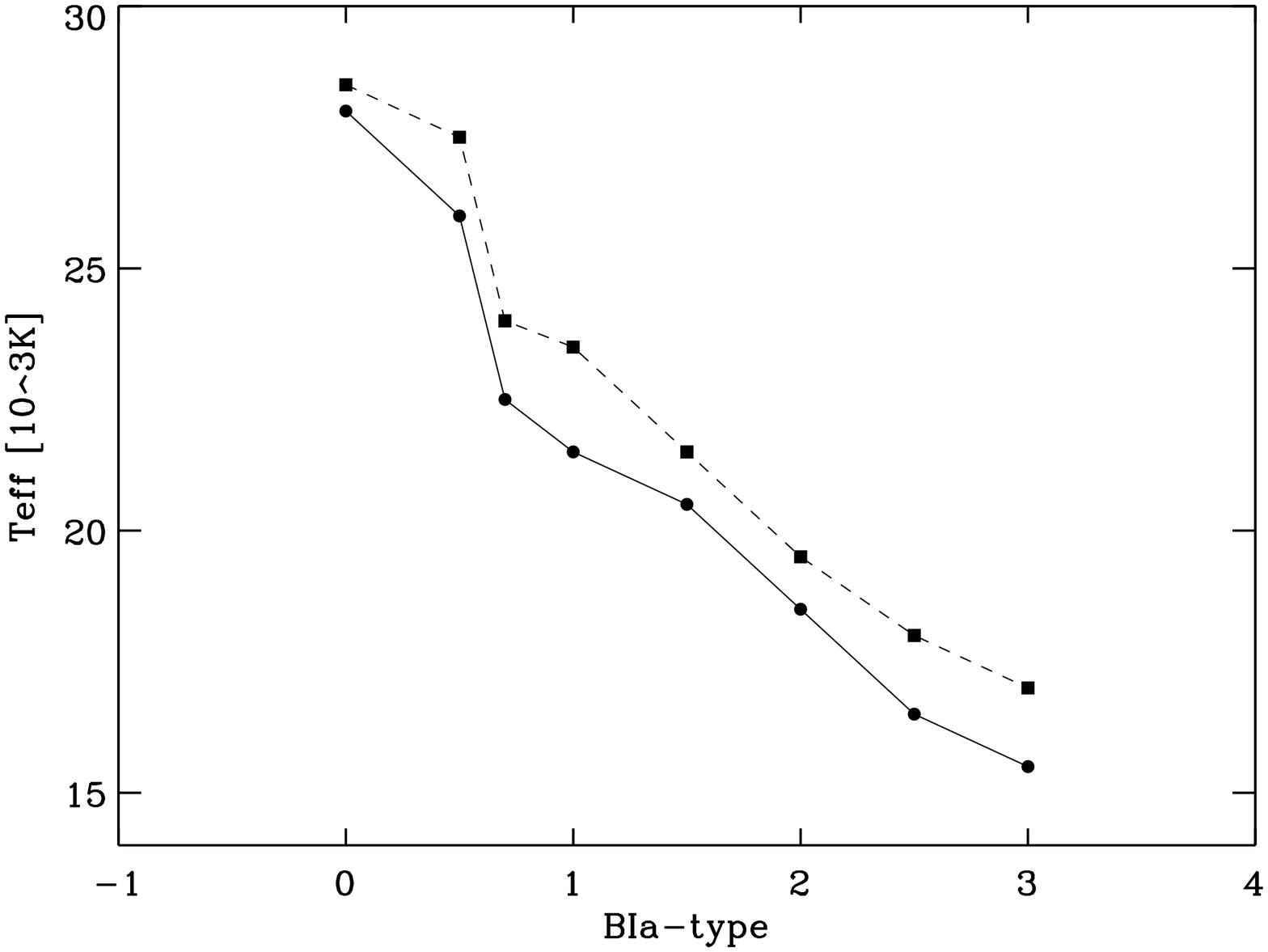,width=7cm,angle=90}
   \psfig{figure=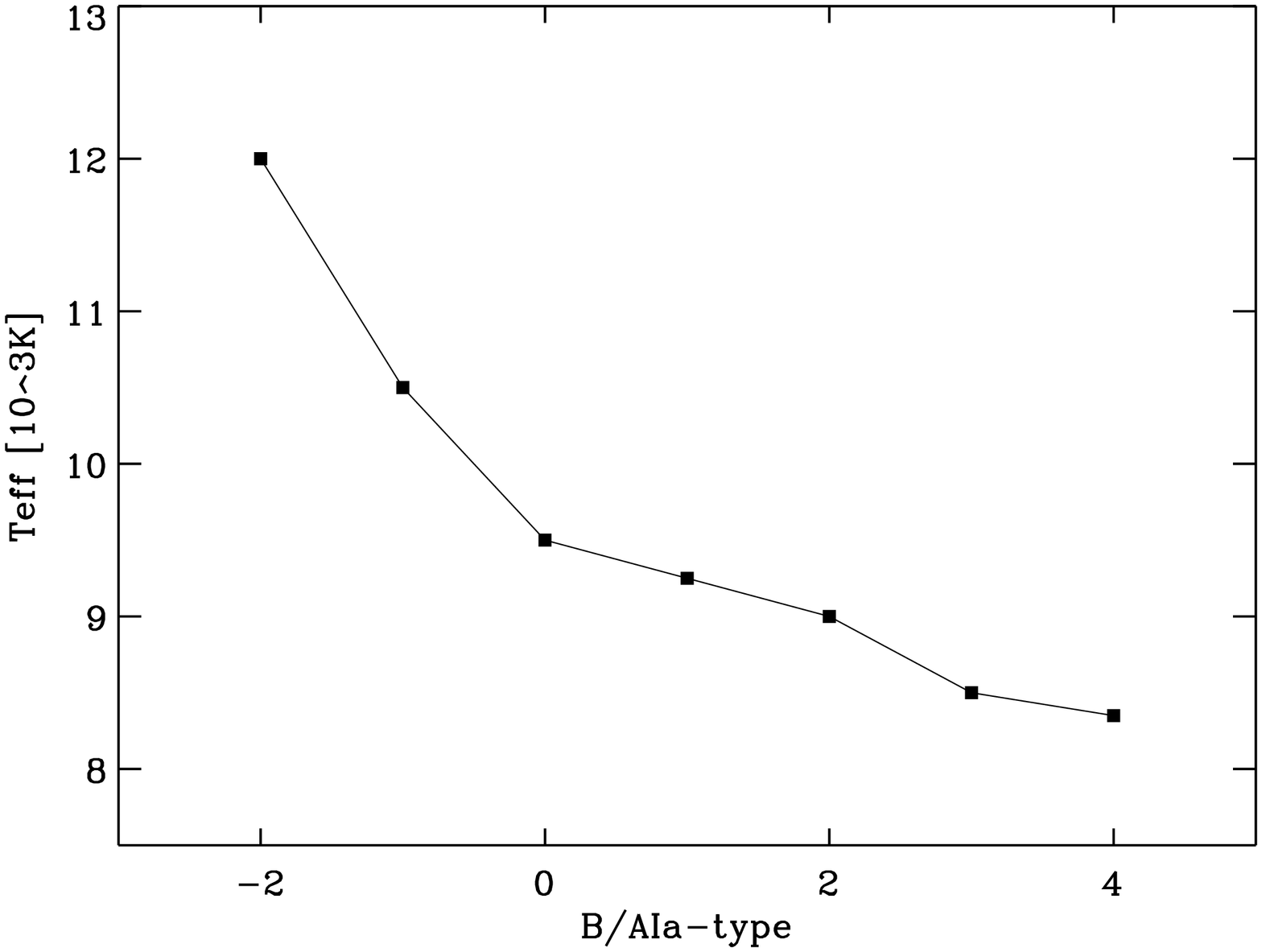,width=7cm,angle=90}
         }}
\caption[]{
The effective temperature scales of early B-supergiants (left, the dashed
curve shows the old scale based on unblanketed NLTE models) and late B and
early A-supergiants (right) of luminosity class Ia. Note that in the right
plot on the abcissa the value of -2 corresponds to spectral type B8.
\label{bateff}}
\end{figure}

As shown in Fig.~\ref{hrd} massive O-stars evolve into B and later A
supergiants. Based on unblanketed, hydrostatic NLTE model atmospheres an
effective temperature scale was introduced for galactic B
supergiants by McErlean et al. (1998, 1999). More recently, \cite{trundle04},
\cite{trundle05}, and \cite{crowther05} have analyzed a large sample of 
Milky Way and SMC B supergiants with the improved models and have revised the
effective
temperature scale for the earliest spectral types of luminosity class Ia.
The results are shown in Fig.~\ref{bateff}. Note that if the spectral
classification scheme introduced by \cite{lennon97} is used, the effective
temperature scale should not depend on metallicity. This is indeed the case,
as a
comparison of the SMC sample (\cite{trundle05}) with the MW sample
(\cite{crowther05}) shows.

\cite{kud03c} have recently introduced an effective temperature scale for
late B and early A supergiants of luminosity class Ia and of solar metallicity,
which is also displayed in Fig.~\ref{bateff}. A discussion of the
metallicity dependence can be found in \cite{evans03}.

\section{Extragalactic stellar astronomy with B and A supergiants}

As shown in Fig.~\ref{hrd} O-stars with masses between 15 to 40\,
M$_{\odot}$ evolve into supergiants of spectral B and A. Although not as
massive and luminous as the most massive O-stars, these stars are the 
brightest ``normal'' stars at
visual light with absolute magnitudes $-7.0  \ge M_{V} \ge -9.5$,
see \cite{bres03}. (By ``normal'' we mean stars evolving peacefully without 
showing signs of eruptions or explosions, which are difficult to handle 
theoretically and observationally). While O-stars emit most of their 
radiation in the extreme and far UV because of their high atmospheric
temperature, late B and early A supergiants are cooler and their
bolometric corrections are much smaller because of Wien's law 
so that their brightness at visual light 
reaches a maximum value during stellar evolution. It is this enormous intrinsic
brightness at visual light, which makes them extremely interesting 
for extragalactic studies far beyond the Local Group.

During their  smooth evolution from the left to the right in
the HRD massive stars are crossing the temperature range of late B and early
A-supergiants in a timescale on
the order of several $10^{3}$ years (\cite{meynet00}). During this short 
evolutionary phase stellar winds with mass-loss rates of the order 
$10^{-6}$ M$_{\odot}$\,yr$^{-1}$ or less
(\cite{kupu00}) do not have enough time to reduce the mass of the star 
significantly so that the mass remains constant. In addition, as 
Fig.~\ref{hrd} shows, the luminosity
stays constant as well. The fact that the evolution of these objects can 
very simply be described by constant mass, luminosity and a straightforward
mass-luminosity relationship makes them a very attractive stellar distance
indicator, as we will explain later in this review.

As evolved objects the blue supergiants are older than their O-star
progenitors, with ages between 0.5 to 1.3\,$\times$\,$10^{7}$ years 
(\cite{meynet00}). 
All galaxies with ongoing star formation or bursts of this age will show
such a population. Because of their age they are spatially less 
concentrated
around their place of birth than O-stars and can frequently be found as 
isolated field stars. This together with their intrinsic brightness makes them
less vulnerable as distance indicators against the effects of crowding even at
larger distances, where less luminous objects such as Cepheids and RR\,Lyrae
start to have problems.

With regard to the crowding problem we also note that the short evolutionary
time of $10^{3}$ years makes it generally very unlikely that an unresolved
blend of two supergiants with very similar spectral types is observed. On
the other hand, since we are dealing with spectroscopic distance indicators,
any contribution of unresolved additional objects of different spectral type
is detected immediately, as soon as it affects the total magnitude 
significantly.

Thus, it is very obvious that blue supergiants seem to be ideal to investigate
the properties of young populations in galaxies. They can be used to study
reddening laws and extinction, detailed chemical composition, i.e. not only
abundance patterns but also gradients of abundance patterns as a function of
galactocentric distance, the properties of stellar winds as function of
chemical composition and the evolution of stars in different galactic
environment. Most importantly, as we will demonstrate below, they are
excellent distance indicators.

\section{Quantitative stellar spectroscopy beyond the Local Group}

Enormous progress has been made over the last years in the development of 
accurate NLTE spectral diagnostics of B and A supergiants. Using high
resolution and high S/N spectra \cite{urba04}, \cite {urba05b},
\cite{trundle04}, \cite{trundle05}, and \cite{crowther05} have studied early
B-supergiants in Local Group galaxies and the Milky Way to determine stellar
properties, chemical composition and abundance gradients. For late B and
early A supergiants Przybilla et al. (2001a,b,c, 2006) have developed very
detailed model atoms for the NLTE radiative transfer diagnostics, which
allow for an extremely accurate determination of effective temperature ($1\%$),
gravities (0.05 dex), and chemical abundances (0.1 dex). Detailed abundance
studies of A supergiants in many Local Group galaxies using these NLTE methods 
were carried out by Venn et al. (1999,
2000, 2001, 2003) and \cite{kaufer04}. The analysis technique is
similar to O-stars, except that different ionization equilibria are used for
the determination of effective temperatures (SiII/III/IV for early B
supergiants and OI/II, NI/II, MgI/II, SII/III for late B and early A
supergiants, see \cite{kud03b} for a more detailed description).

\subsection{Chemical composition}

For extragalactic applications beyond the Local Group spectral resolution
becomes an issue. The important points are the following. Unlike the case
of late type stars, crowding and blending of lines is not a severe problem
for hot massive stars, as long as we restrict our investigation to the visual
part of the spectrum. In addition, it is important to realize that massive
stars have angular momentum, which leads to usually high rotational velocities.
Even for A-supergiants, which have already expanded their radius
considerably during their evolution and, thus, have slowed down their
rotation, the observed projected rotational velocities are still on the order
of 30\,km\,s$^{-1}$ or higher. This means that the intrinsic full half-widths 
of metal lines are on the order of 1\,{\AA}. In consequence, for the detailed
studies of supergiants in the Local Group a resolution of 25,000 sampling a 
line with five data points is ideal. This is indeed the resolution, which has
been applied in most of the work referred to above.

\begin{figure}
\centerline{\hbox{
   \psfig{figure=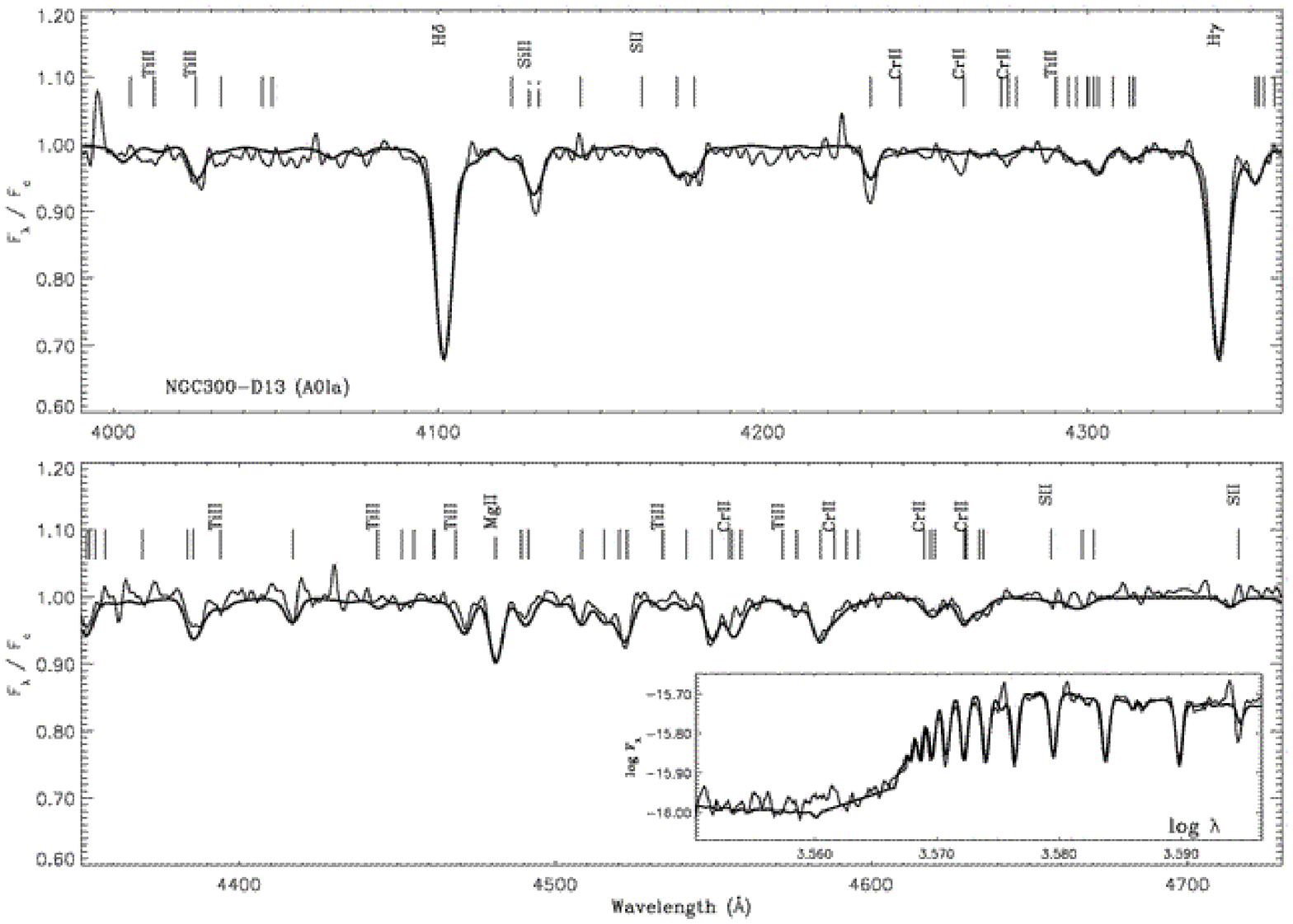,width=12cm,angle=90}
         }}
\caption[]{
Quantititative spectral analysis of an A-supergiant in the Sculptor spiral
galaxy NGC 300 at a distance of 2 Mpc. The inset shows the fit of the Balmer
jump, which is used to determine T$_{eff}$. (From \cite{kud06b}, see also
\cite{bres02a}).
\label{specngc300a}}
\end{figure}

\begin{figure}
\centerline{\hbox{
   \psfig{figure=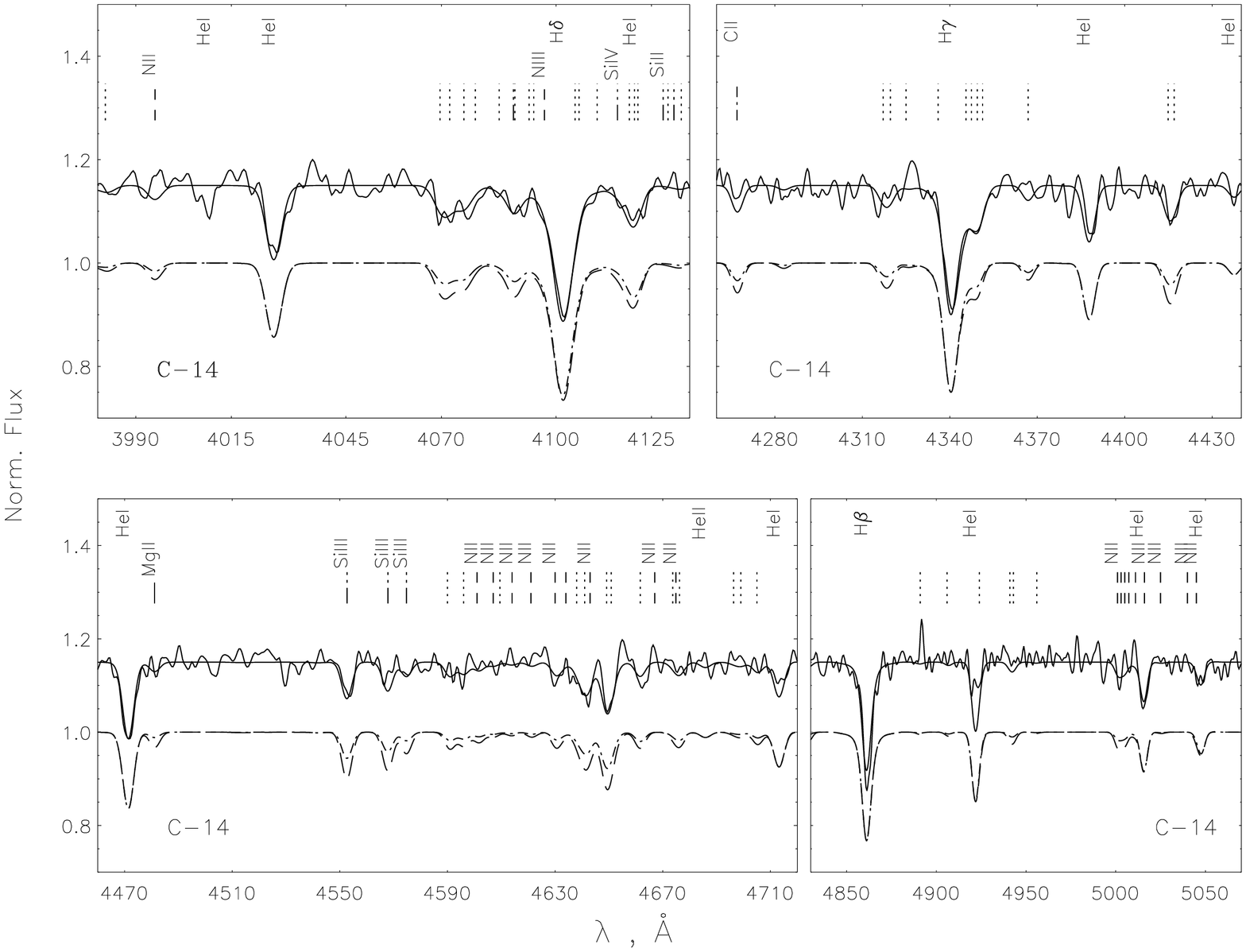,width=12cm,angle=90}
         }}
\caption[]{
Quantititative spectral analysis of an early B-supergiant in the Sculptor spiral
galaxy NGC 300. (From \cite{urba05a}).
\label{specngc300b}}
\end{figure}

However, as we have found out empirically (\cite{przy02}), degrading the
resolution to 5,000 (FWHM = 1\,{\AA}) has only a small effect on the accuracy
of the diagnostics, as long as the S/N remains high (i.e. 50 or better).
Even for a resolution of 2,500 (FWHM = 2\,{\AA}) it is still possible to
determine T$_{eff}$ to an accuracy of 2 percent, $\log g$ to 0.05 dex and
individual element abundances to 0.1. 

\begin{figure}
\centerline{\hbox{
   \psfig{figure=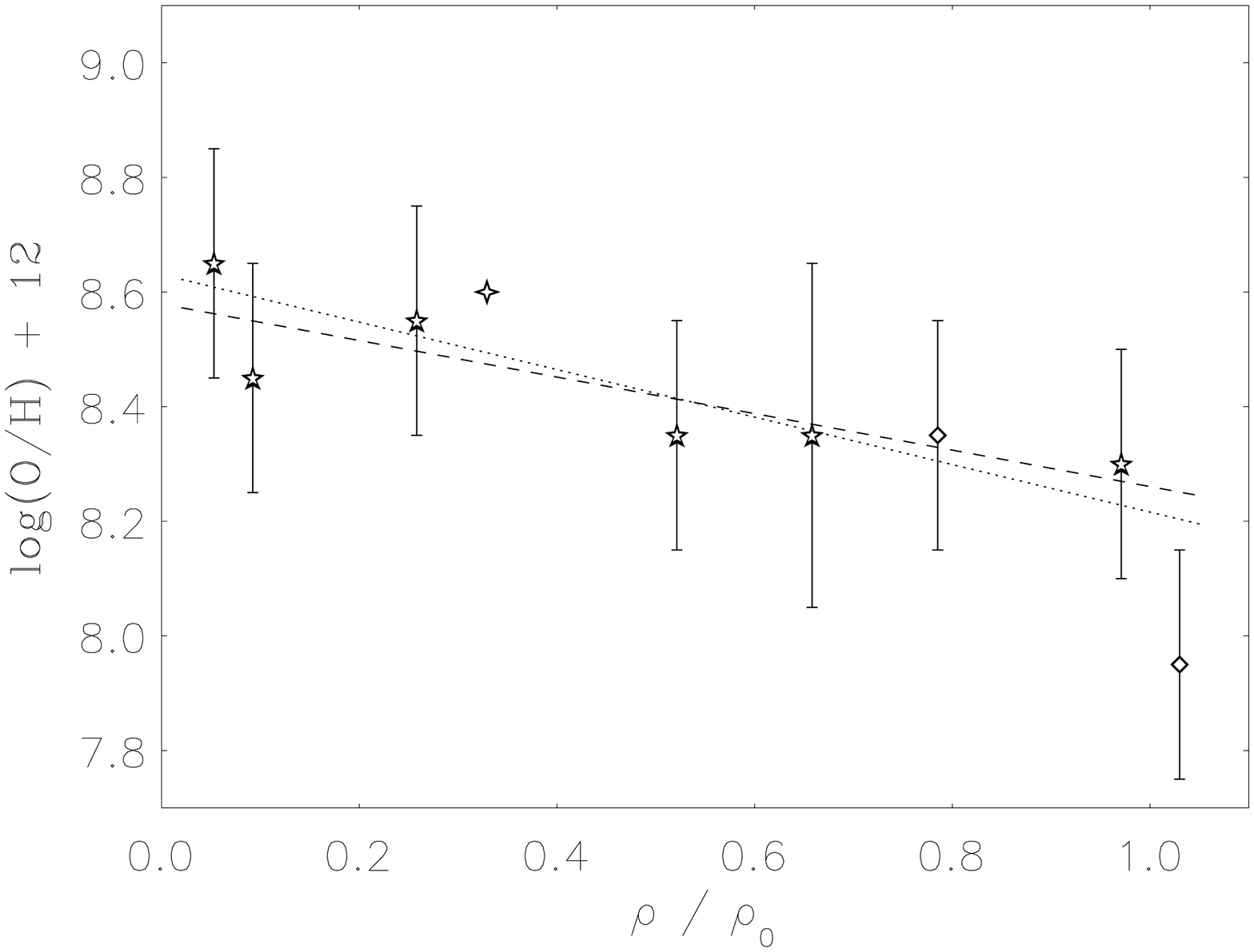,width=7cm,angle=90}
   \psfig{figure=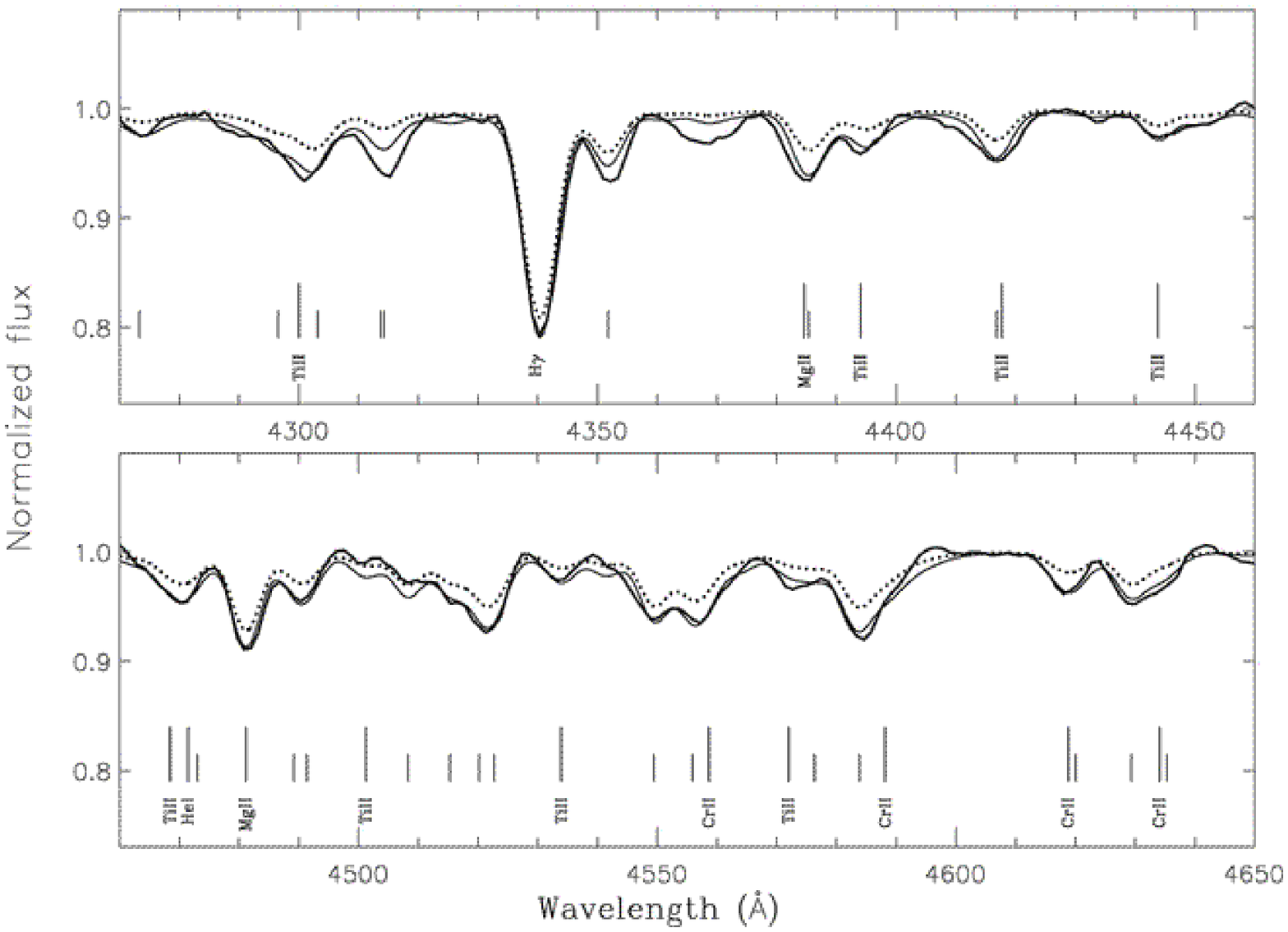,width=7cm,angle=90}
         }}
\caption[]{
Left: the metallicity gradient of the young stellar population in the 
Sculptor
spiral galaxy NGC 300 as detemined for quantitative stellar spectroscopy. 
(From \cite{urba05a}, Bresolin et al. (2002a,b). Right: Quantitative 
spectral analysis of an A-supergiant in the spiral
galaxy NGC 3621 at a distance of 7 Mpc. (From \cite{bres01}).
\label{abundngc300}}
\end{figure}

Bresolin et al. (2001, 2002a, 2002b, 2003, 2006), Urbaneja et al. (2003,
2005a, 2006), \cite{kud03c} and \cite{kud06b} have used FORS at the VLT 
with a resolution of 1,000 
(FWHM = 5\,{\AA}) to study blue supergiants far beyond the Local Group. The
accuracy in the determination of stellar properties at this rather low
resolution is still remarkable. The effective temperature (for late B and A
supergiants now determined
from the Balmer jump rather than from ionization equilibria) is accurate to
roughly 4 percent and the determination of gravity based on fitting the broad
Balmer lines remains unaffected by the lower resolution and is still 
good to 0.05\,dex. Abundances
can be determined with an accuracy of 0.2\,dex. In Fig.~\ref{specngc300a}, 
\ref{specngc300b}, and \ref{abundngc300} we show examples of the detailed 
spectral fits that can be accomplished. Fig.~\ref{abundngc300} demonstrates,
how important information about the metallicity gradients of the young
stellar population in spiral galaxies can be obtained directly from the
spectral analysis of blue supergiants.

\subsection{Extragalactic distance determinations with the ``Flux Weighted
Gravity - Luminosity Relationship (FGLR)''}

The best established stellar distance indicators, Cepheids and RR\,Lyrae,
suffer from two major problems, extinction and metallicity dependence, both 
of which
are difficult to determine for these objects with sufficient precision. Thus,
in order to improve distance determinations in the local universe and to
assess the influence of systematic errors there is
definitely a need for alternative distance indicators, which are at least as
accurate but are not affected by uncertainties arising from extinction or
metallicity. Blue supergiants are ideal objects for 
this purpose because of their enormous intrinsic brightness, which makes 
them available for accurate quantitative 
spectroscopic studies even far beyond the Local Group using the new generation
of 8m-class telescopes and the extremely efficient multi-object spectrographs 
attached to them (see previous subsection). 
Quantitative spectroscopy allows us to
determine the stellar parameters and thus the intrinsic energy distribution,
which can then be used to measure reddening and the extinction law. In
addition, metallicity can be derived from the spectra. We emphasize that a 
reliable {\em spectroscopic}
distance indicator will always be superior, since an enormous amount of
additional information comes for free, as soon as one is able to obtain a
reasonable spectrum.

\begin{figure}
\centerline{\hbox{
   \psfig{figure=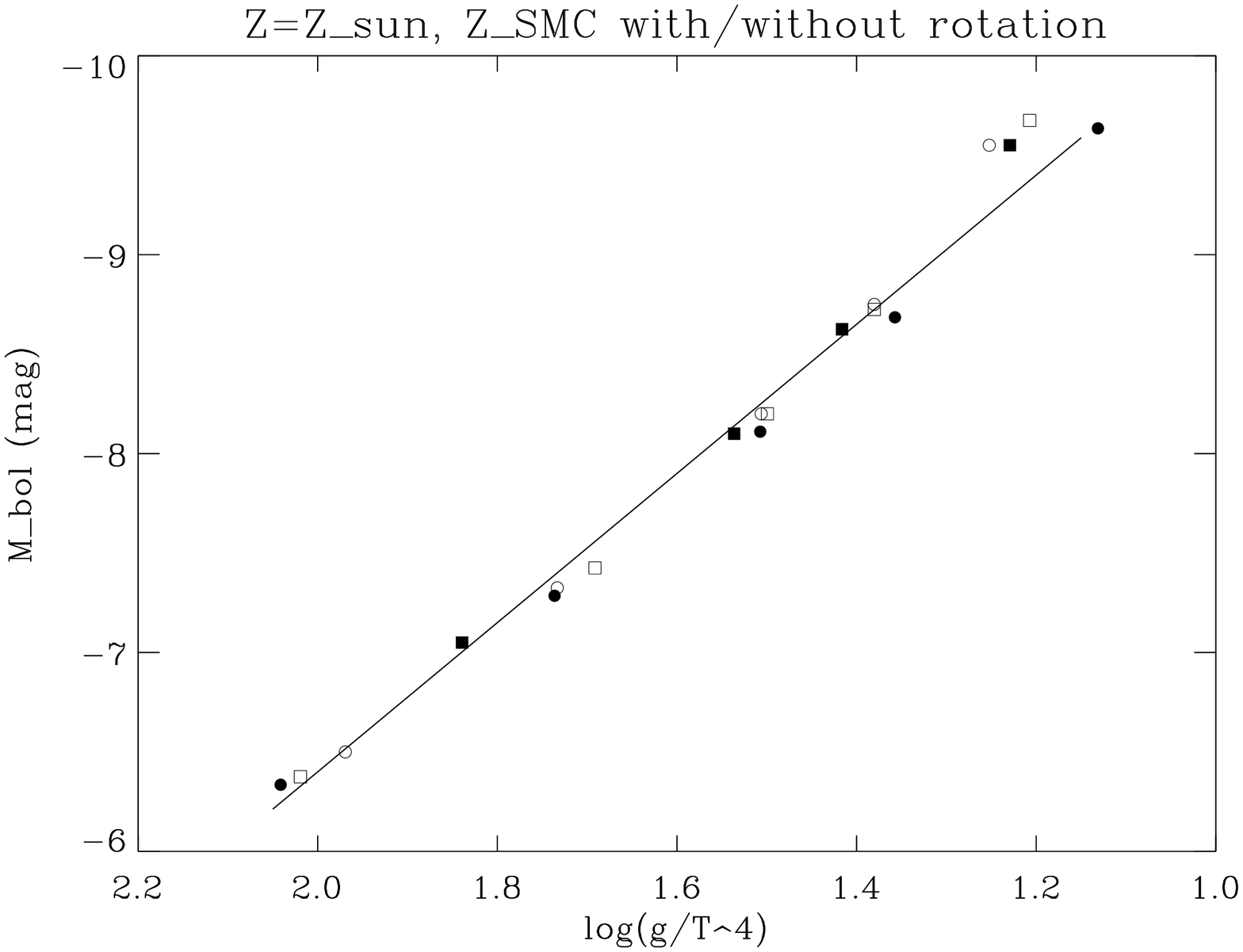,width=7cm,angle=90}
   \psfig{figure=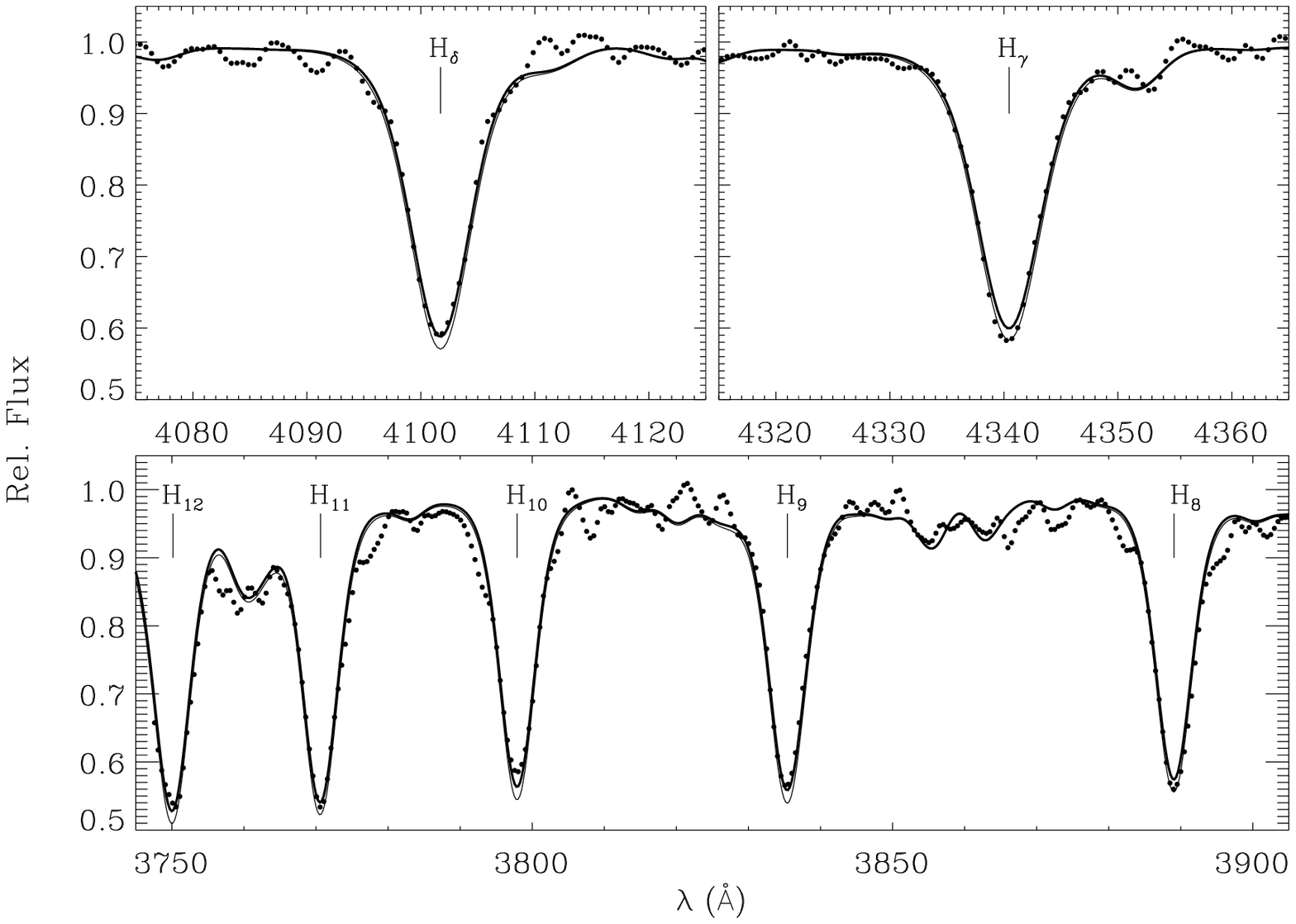,width=7cm}
         }}
\caption[]{
Left: The FGLR of stellar evolution models from \cite{meynet94} and~\cite{meynet00}.
Circles correspond to models with rotation, squares represent models without
the effects of rotation. Solid symbols refer to galactic metallicity and
open symbols represent SMC metallicity. The solid curve corresponds to 
$a=-3.75$ in the FGLR. Note that gravities are in cgs units and the
temperatures are in units of 10$^{4}$K.
Right: Fit of the higher Balmer lines of an A-supergiant in the Sculptor 
galaxy 
NGC\,300 using two atmospheric models with T$_{eff}=9500$\,K and 
$\log g=1.60$ ({\it thick line}) and 1.65 ({\it thin line}), respectively. 
The data were taken with
FORS at the VLT. For further discussion see \cite{kud03c}
\label{fglrev}}
\end{figure}

A very promising spectroscopic distance determination method based on simple
stellar physics is the \emph{Flux-Weighted Gravity -- Luminosity
Relationship} (FGLR), which was introduced by~\cite{kud03c}. 
When discussing Fig.~\ref{hrd} in Sect.~1 we noted
that massive stars evolve through the domain of blue supergiants with
constant luminosity and constant mass. This has a very simple, but very
important consequence for the relationship of gravity and effective
temperature along each evolutionary track. From
\begin{equation}
L \propto R^{2}T^{4}_\mathrm{eff} = \mathrm{const.} ; M = \mathrm{const.}
\end{equation}
follows immediately that
\begin{equation}
M \propto g\;R^{2} \propto L\;(g/T^{4}_\mathrm{eff}) = \mathrm{const.}
\end{equation}

This means that each object of a certain initial mass on the ZAMS
has its specific value of the \emph{``flux-weighted gravity''
g/T$^{4}_{eff}$} during the blue supergiant stage. 
This value is 
determined by the relationship between stellar mass and luminosity, which to
a good approximation is a power law
\begin{equation}
L \propto M^{x}\;.
\end{equation}

Inspection of evolutionary calculations with mass-loss, cf. \cite{meynet94}
and \cite{meynet00}, shows that $x=3$ is a good value in the range of
luminosities considered, although $x$ changes towards higher masses.
With the mass -- luminosity power law we then obtain
\begin{equation}
L^{1-x} \propto (g/T^{4}_\mathrm{eff})^{x}\;,
\end{equation}
or with the definition of bolometric magnitude 
$M_\mathrm{bol}$\,$\propto$\,$-2.5\log L$
\begin{equation}
-M_\mathrm{bol} = a\log(g/T_{\!\mbox{\scriptsize eff}}^4) + b\;.
\end{equation}

This is the FGLR of blue supergiants. Note that the proportionality constant
$a$ is given by the exponent of the mass -- luminosity power law through
\begin{equation}
a = 2.5 x/(1-x)\;.
\end{equation}
and $a=-3.75$ for $x=3$. Mass-loss will depend on metallicity and
therefore affect the mass -- luminosity relation. In addition, stellar rotation
through enhanced turbulent mixing might be important for
this relation. In order to investigate these effects we have used the models of
\cite{meynet94} and~\cite{meynet00} to construct the stellar evolution FGLR, 
which is displayed in Fig.~\ref{fglrev}. The result is very encouraging. All
different models with or without rotation and with significantly different
metallicity form a well defined very narrow FGLR.

\begin{figure}
\centerline{\hbox{
   \psfig{figure=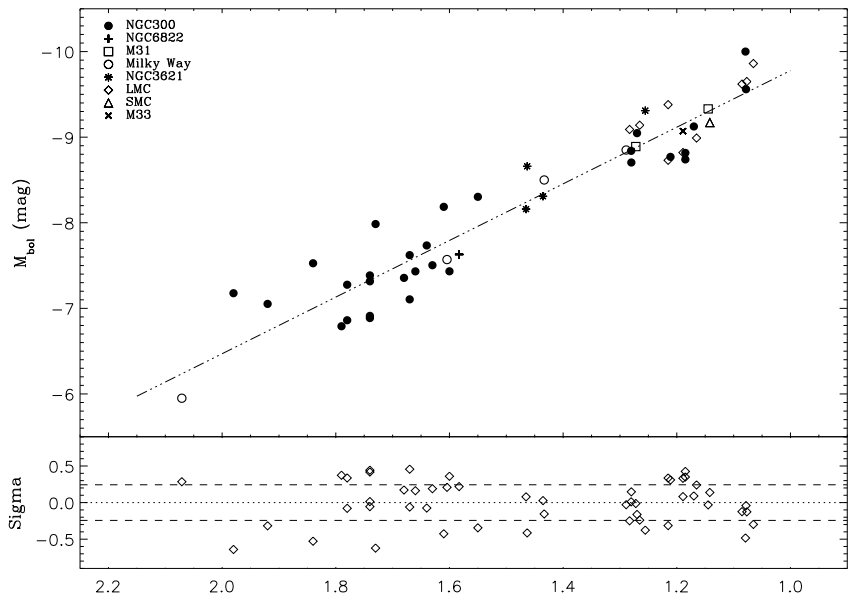,width=12cm,angle=90}
         }}
\caption[]{
The FGLR of B and A supergiants in Local Group galaxies and in the spiral
galaxies NGC\,300 and NGC\,3621 at a distance of 2 and 7 Mpc, respectively.
The abcissa is the same as in Fig.~\ref{fglrev} (left).
(From \cite{kud06b}, see also \cite{kud03c}). For a discussion, see text.
\label{fglr1}}
\end{figure}

In order to verify the existence of the theoretically predicted FGLR
\cite{kud03c} re-analyzed a large sample of late B and early A supergiants in
several Local Group galaxies (high resolution spectra) and in the spiral
galaxies NGC 300 (2Mpc) and NGC 3621 (7Mpc) using VLT FORS low resolution
spectra as descibed above. Fig.~\ref{fglrev} demonstrates how precisely the
gravities can be determined even at low resolution. The effective
temperatures were obtained from the spectral types using the relation
displayed in Fig.~\ref{bateff}. Taking into account the valid criticism by
\cite{evans03} about the metallicity dependence of the temperature vs.
spectral type relationship, \cite{kud06b} used the
information about the Balmer jump in the VLT/FORS spectrophotometric data (see
Fig.~\ref{specngc300a}) to determine effective temperature and metallicity
independently. This new procedure lead to the FGLR shown in Fig.~\ref{fglr1}.
A least square fit yields a=-3.31 and b=13.09 with a one $\sigma$ scatter of
0.24 mag. Fixing the slope to the theoretical value a=-3.75 we obtain as a zero point 
b=-13.73 with $\sigma$=0.25. 

The FGLR as displayed in Fig.~\ref{fglr1} is
an extremely tight relationship with a scatter comparable to the observed 
period-luminosity relationships of Cepheids. We conclude that blue supergiants
provide a great potential as excellent
extragalactic distance indicators. The quantitative analysis of their spectra
-- even at only moderate resolution -- allows the determination of stellar 
parameters, stellar wind properties and chemical composition with remarkable
precision. In addition, since the spectral analysis yields intrinsic energy
distributions over the whole spectrum from the UV to the IR, multi-colour
photometry can be used to determine reddening, extinction laws and extinction.
This is a great advantage over classical distance indicators, for which only
limited photometric information is available, when observed outside the Local
Group. Spectroscopy also allows to deal with the effects of crowding and
multiplicity, as blue supergiants, due to their enormous brightness,
are less affected by such problems than for instance Cepheids, which are
fainter.

Applying the FGLR method on objects brighter than
$M_{V}=-8$\,mag and using multi-object spectrographs at 8 to 10m-class
telescopes, which allow for quantitative spectroscopy down to
$m_{V}=22$\,mag, we estimate that with 20 objects per galaxy we will be
able to determine distances out to distance moduli of $m-M \sim30$\,mag
with an accuracy of 0.1\,mag. We emphasize that these distances will not be 
affected by uncertainties in extinction and metallicity, because we will be 
able to derive the corresponding quantities from the spectrum.

\section{Winds of hot massive stars}

All hot massive stars have winds, which are driven by radiation. As
emphasized in the introduction these winds are fundamentally important for
the spectral diagnostics of massive stars, for their evolution and for the
galactic environment. Over the last decades very detailed and refined 
methods have been developed for the diagnostic of stellar winds and for
modelling their hydrodynamics. A comprehensive review describing these methods
and summarizing the basic properties of stellar winds was published
a few years ago by \cite{kupu00} (see also \cite{kud00}, which was published
in these Symposium Series). The mass-loss rates and terminal velocities of
these winds are related to the physical parameters of massive hot stars
through simple relationships. The stellar wind momentum $\dot{M}v_{\infty}$
is related to radius R, luminosity L and metallicity Z through

\begin{equation}
\dot{M} v_{\infty} R^{1/2} \propto L^{1.8} (Z/Z_{\odot})^{-0.8}
  \label{mom}
\end{equation}

This the wind momentum luminosity relationship (WLR), which has been 
introduced first by \cite{kud95}. (Note that the left hand side is
called the ``modified stellar wind momentum'').
The physics leading to this relationship has been described in detail in
\cite{kud98} and \cite{kud00}. Proportionality constants for the WLR have been
determined by \cite{puls96} for O-stars and by \cite{kud99} for B and A
supergiants. (See also \cite{kupu00}).

The terminal velocities $v_{\infty}$ of the winds of hot massive stars are 
related to the photospheric escape velocities $v^{phot}_{esc}$ and metallicity
through

\begin{equation}
v_{\infty} \propto v^{phot}_{esc} (Z/Z_{\odot})^{-0.15}
  \label{vel}
\end{equation}
\begin{equation}
v^{phot}_{esc} = (2GM(1-\Gamma)R^{-1})^{1/2}
\end{equation}

The proportionality constant in Eq.~\ref{vel} is about 2.6 ($\pm 0.5$
)  for O-stars and
early B-supergiants but becomes smaller for lower temperatures and is about
1.0 for A-supergiants (see \cite{kupu00}, but see also the new results 
discussed below). $\Gamma$ is the usual distance to
the Eddington limit.

\subsection{O-star wind momenta based on new diagnostics 
           with metal line blanketed model atmospheres}

The proportionality constants for the WLR were based on spectral
diagnostics of H$_{\alpha}$, which is very sensitive to the strengths of
stellar winds and, therefore, regarded as a very good indicator of mass-los
rates (see \cite{kupu00}). However, the original work 
was based on the use of NLTE
model atmospheres, which neglected metal line blanketing. It was, therefore,
crucially important to repeat the H$_{\alpha}$ studies of stellar winds with
the improved model atmospheres described in the sections before.

\cite{repo04} and \cite{markova04} carried out a comprehensive study of
O-stars in the Milky Way. Fig.~\ref{odiag1} indicates the general trend
with regard to the diagnostics of wind momenta. Since the analysis with
metal line blanketed NLTE atmospheres yield lower temperatures and, therefore,
lower luminosities, the WLR is simply shifted towards lower luminosities.
Fig.~\ref{mom} shows the observed WLR for O-stars. As in \cite{kupu00} we 
include the
results of similar diagnostics of Central Stars of Planetary Nebulae (CSPN)
(for details see \cite{kud06a}) to demonstrate that very obviously the WLR
extends to these low mass hot stars in a post-AGB evolutionary stage as well. 
\cite{repo04} and \cite{markova04} discuss the difference between the WLR of
dwarfs and supergiants and argue that the theory cannot reproduce this
difference. They conclude that the H$_{\alpha}$ mass-loss rates of
supergiants are affected by inhomogeneous stellar wind clumping and in
reality are probably close to the ones of dwarfs. To check this hypothesis, 
we compare with the theoretical wind momenta for O-stars and CSPN calculated
by \cite{kud02}, who used a new theoretical approach for the theory of line
driven winds. The result is also shown in Fig.~\ref{mom}. While we find a clear
offset between the WLRs of dwarfs and supergiants, the theoretical wind 
momenta are a little too small in both cases, which would mean that clumping
might be important for both dwarfs and supergiants. We will discuss the
effects of clumping at the end of this review.

With the new stellar parameters derived with metal line blanketed models
there is also a change in the values of photospheric escape velocities. After
a first check of all the new results the factor relating terminal velocity
with escape velocity appears to be 3.1 rather than 2.6.

\begin{figure}
\centerline{\hbox{
   \psfig{figure=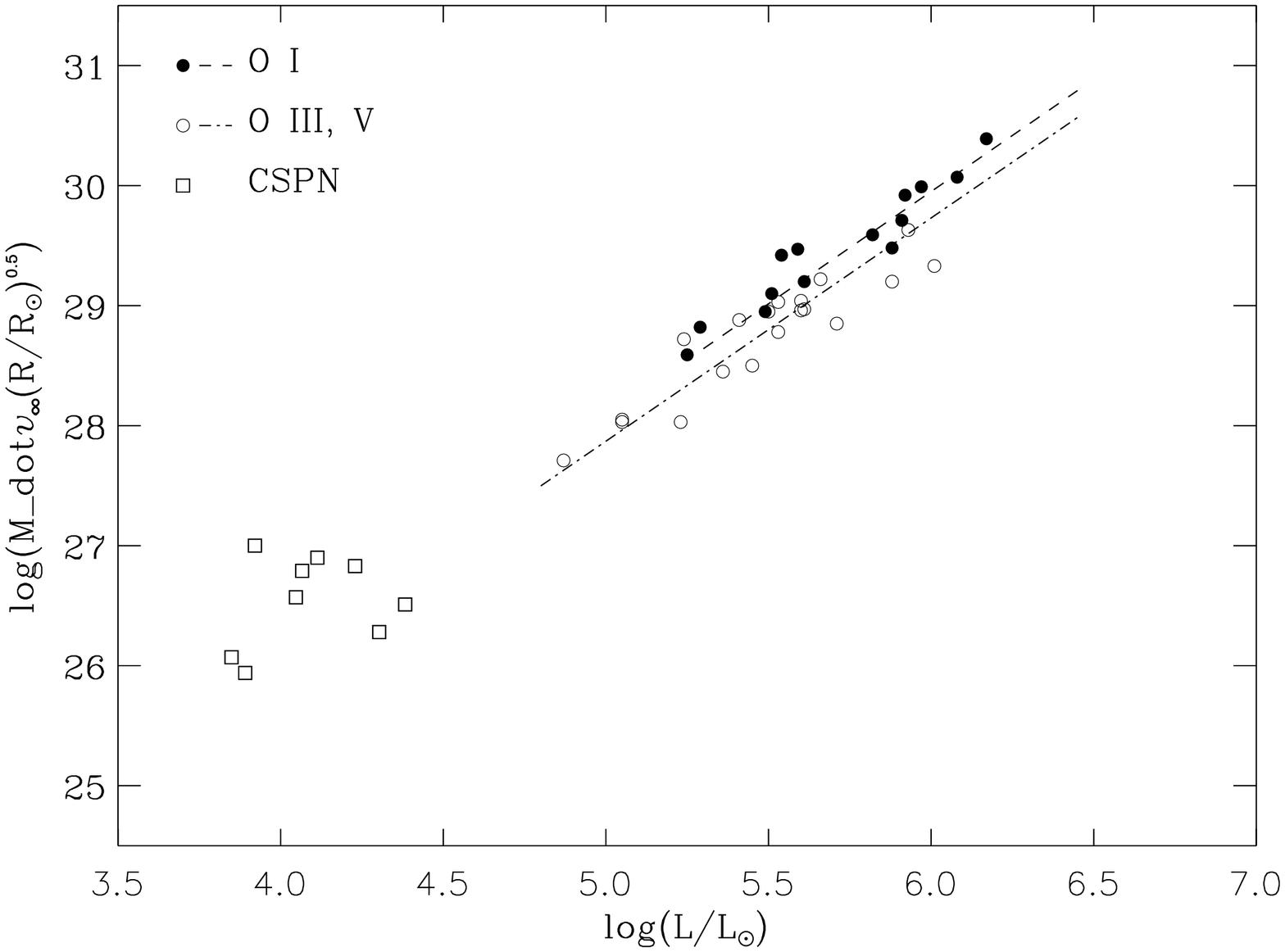,width=7cm,angle=90}
   \psfig{figure=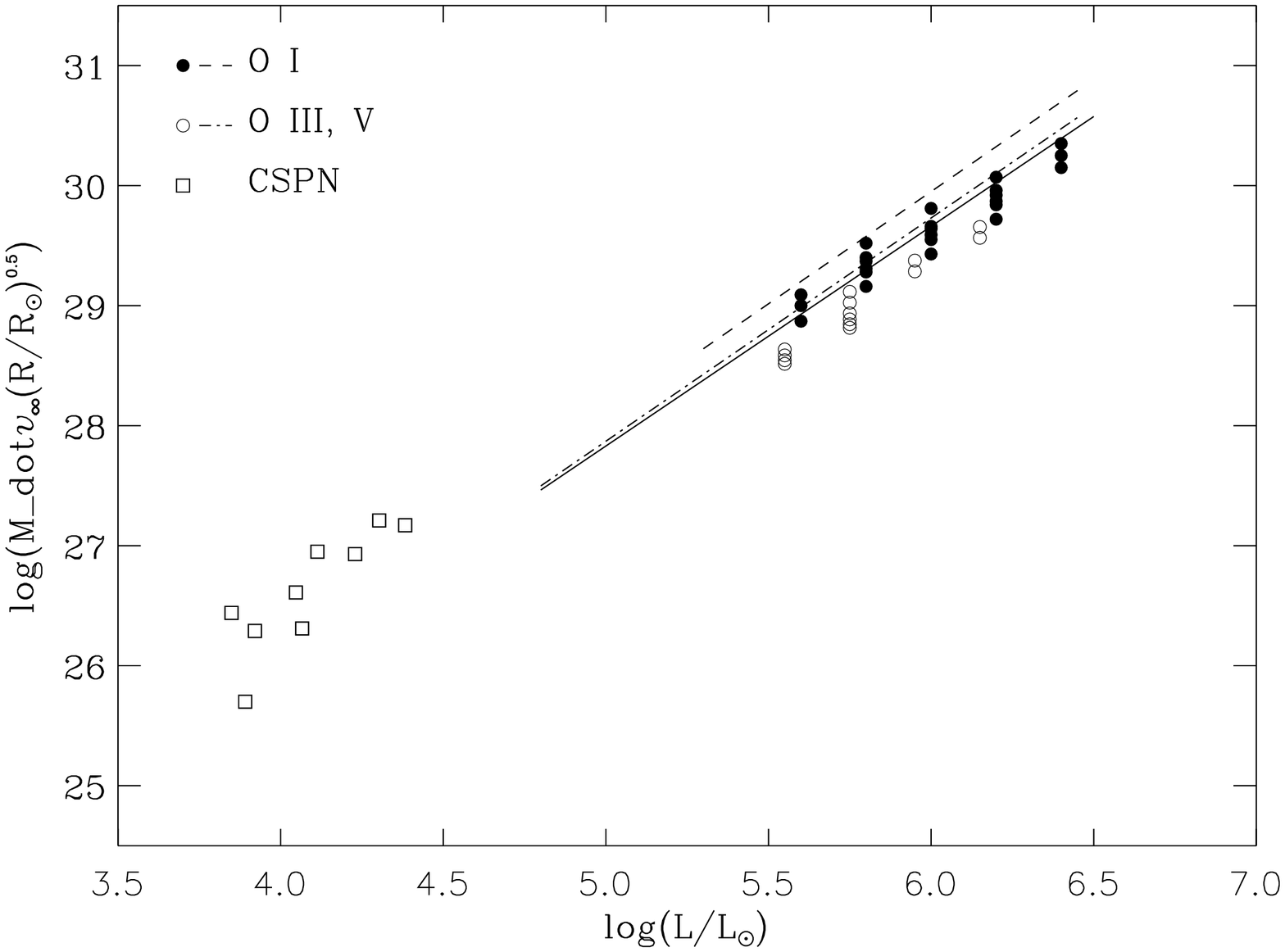,width=7cm,angle=90}
         }}
\caption[]{
Left: Observed wind momenta of O-stars and CSPN as a function of luminosity.
Right: Calculated stellar wind momenta for O-stars and CSPN using the theory 
of line driven winds as developed by \cite{kud02}. Symbols refer to model 
calculations. The dashed lines are the regression curves obtained from the 
observations displayed in the left diagram. The solid line represents the 
theoretical approach by \cite{vink00}.
\label{mom}}
\end{figure}

\subsection{Metallicity dependence}

\begin{figure}
\centerline{\hbox{
   \psfig{figure=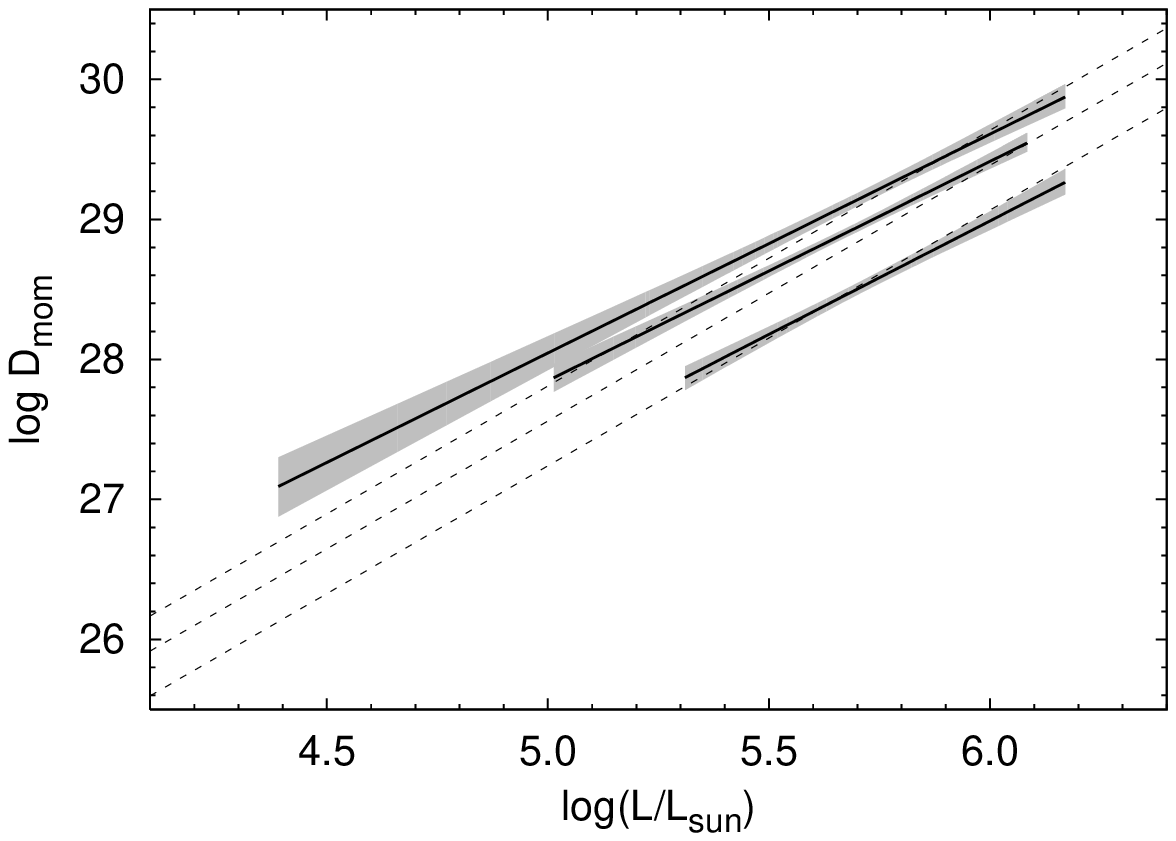,width=7cm}
   \psfig{figure=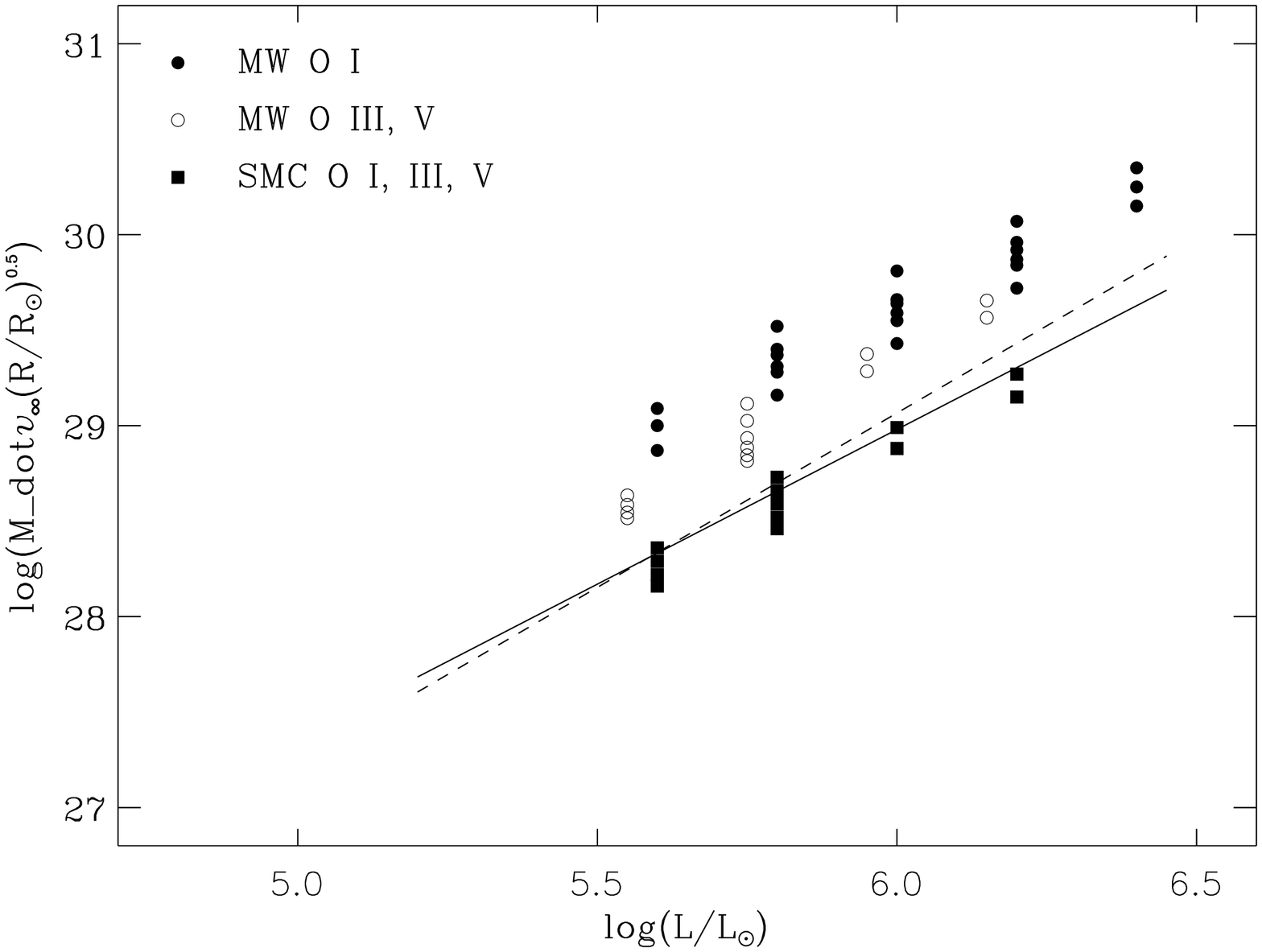,width=7cm,angle=90}
         }}
\caption[]{
Left: Observed wind momenta of O-stars and CSPN as a function of luminosity.
The shaded area represent observational results in the MW, LMC, and SMC and
the solid lines are the observed WLR regression curves. The
dashed curves represent theoretical predictions by \cite{vink01} for MW, LMC
and SMC metallicities, respectively. (From \cite{mokiem06b}).
Right: Calculated stellar wind momenta for O-stars and CSPN using the theory 
of line driven winds as developed by \cite{kud02}. Symbols refer to model 
calculations. The solid curve is the observed regression for the SMC
from the left diagram. The dashed line represents the 
theoretical approach for the SMC by \cite{vink01}.
\label{momsmc}}
\end{figure}

Since the mechanism of driving stellar winds is the absorption of
photospheric photon momentum through many thousands of spectral lines, it is
clear that the strengths of winds is expected to depend on metallicity.
First predictions of the metallicity dependence of both mass-loss rates and
terminal velocities were made by \cite{abbott82} and \cite{kud87}, later
confirmed and extended by \cite{leitherer92}. Improved calculations
were carried out more recently by \cite{vink01} and \cite{kud02}.
Observationally, \cite{puls96} analyzed O-stars in the Clouds relative to
the Milky way and found a first observational indication for the power law
dependence of wind momenta on metallicity. Moreover, \cite{kupu00} showed
that O-stars in the metal poor SMC have lower lower terminal velocities 
than their MW counterparts. More recently, \cite{massey04}, \cite{massey05},
\cite{evans04}, \cite{hillier03}, and \cite{crowther02} have clearly
confirmed these results. Fig.~\ref{momsmc} shows the most recent results
obtained by \cite{mokiem06b}, which indicate excellent agreement with the
predictions of the theory.

\subsection{B supergiants}

\begin{figure}
\centerline{\hbox{
   \psfig{figure=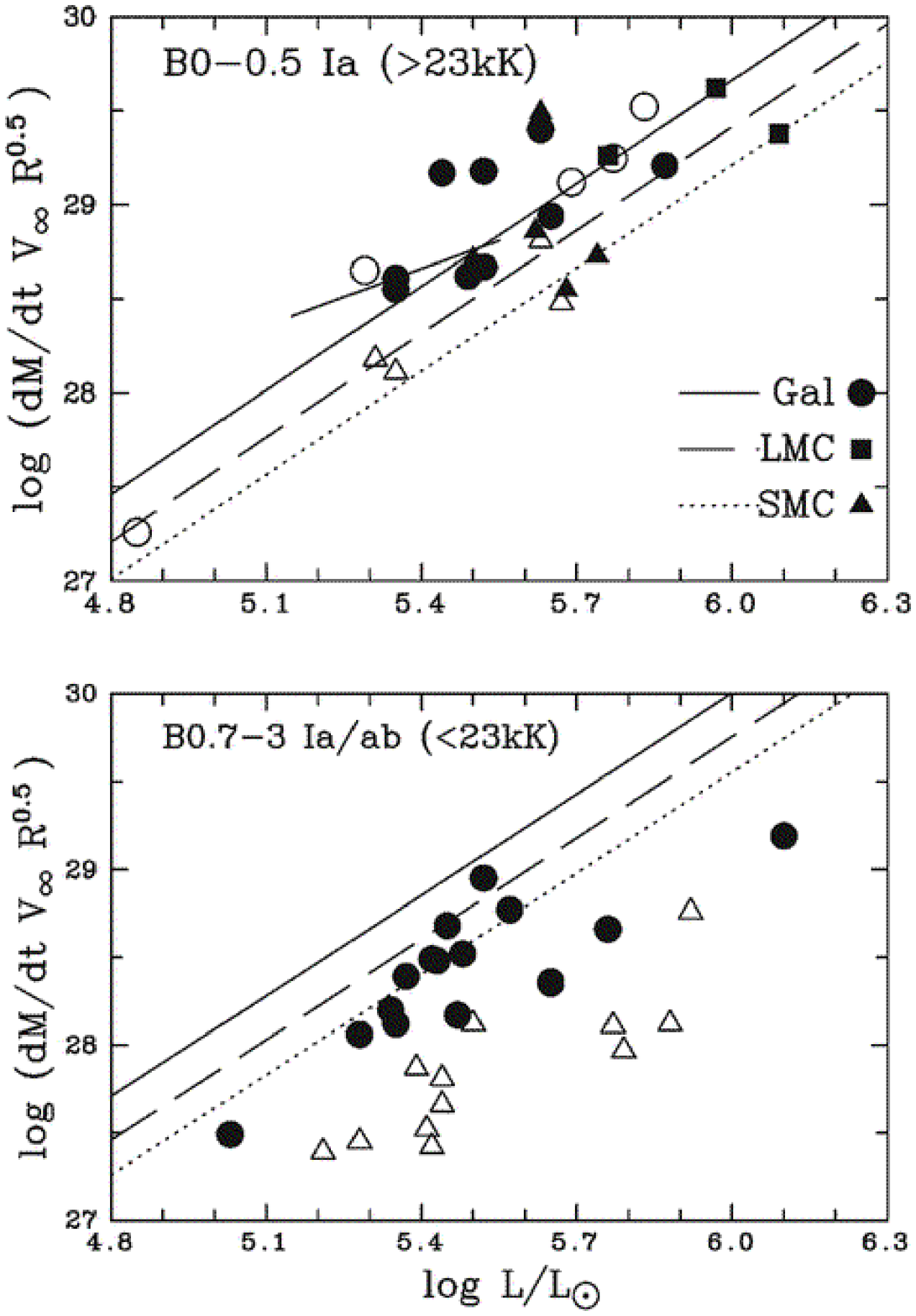,width=7cm}
   \psfig{figure=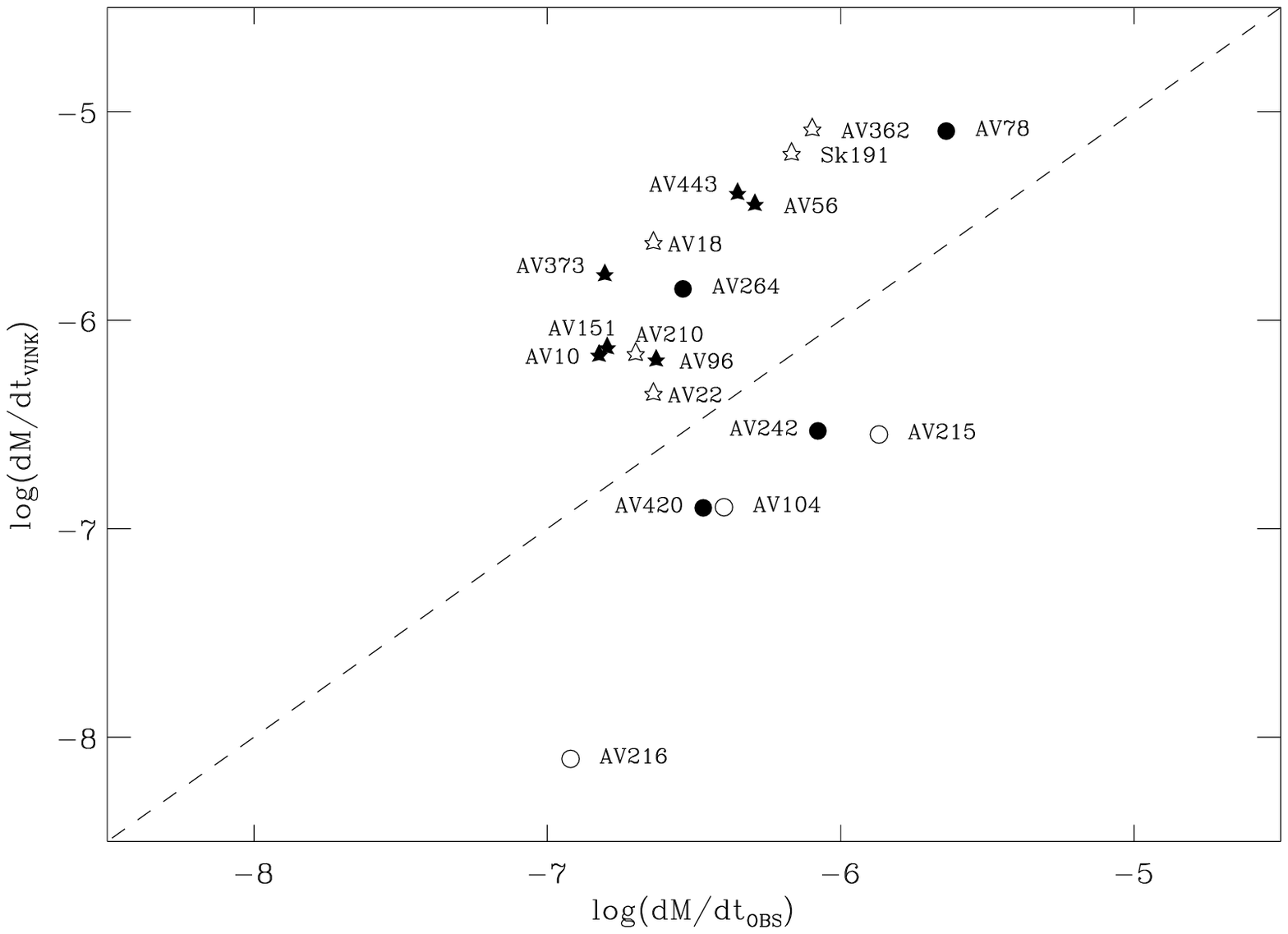,width=7cm}
         }}
\caption[]{
Left: Observed wind momenta of BIa supergiants as a function of luminosity
for objects hotter than 23kK (B0-B0.5) (upper panel) and cooler than 23kK
(B0.7-3) (lower panel) and for objects in the Galaxy(circles), LMC (squares)
and  SMC (triangles). Overplotted are wind momenta predicted by the
computations by Vink et al. (2000, 2001) for the metallicities of the three
galaxies. (from (\cite{crowther05}).
Right: Mass-loss rates as predicted by Vink et al. (2001) for SMC
B-supergiants compared to observations. Circles represent early type
supergiants, stars corresponds to mid types. 
\label{bmom}}
\end{figure}

Because of changes in ionization leading to different sets of spectral lines
absorbing photospheric photon momentum and driving the stellar winds the WLR
is expected to be spectral type dependent. Indeed, \cite{kud99} in their study
of winds of B and A supergiants found a strong variation of the WLR with 
spectral type. While early B supergiants seemed to have wind momenta only
somewhat weaker than their O-stars counterparts, mid B supergiants showed
much weaker winds, whereas A supergiants had stellar wind momenta comparable
to the early B spectral types, but a steeper slope of the WLR.

With the new line blanketed models available \cite{trundle04},
\cite{trundle05}, \cite{urba04}, \cite{evans04} and 
\cite{crowther05} have re-investigated
the wind properties of B-supergiants. In general, the results by \cite{kud99}
are confirmed. There is still on average a difference in the strengths of 
wind momenta between early and mid B-spectral types. However, while mass-loss 
rates for early spectral types are very similar to those obtained with the 
unblanketed models by \cite{kud99}, the new values for mid B-supergiants are 
about a factor of 3 higher on average, but even with this increase they remain
significantly lower than those for the earl B types (see Fig.~\ref{bmom}).
There is a clear metallicity dependence of wind momenta.

As shown in Fig.~\ref{bmom} the theory of line driven winds fails to
reproduce these observations. Vink et al. (2000, 2001) predict strong
increase of mass-loss rates below 24kK due to changes of ionization, which is 
definitely not observed. For the early type supergiants show mass-loss rates 
much stronger than predicted by the theory. While this can be explained by
stellar wind clumping, we believe that that in general more theoretical work
for B-supergiant winds is needed.

The ratio between terminal and escape velocities changes as a function of 
effective temperature showing a similar trend as found by \cite{prinja98}, 
but quantitatively different. For temperatures above 24kK the ratio is 3.4,
between 20kK and 24kK it is 2.5 and below 20kK  1.9 is
found. In all temperature ranges there is a large scatter around these average
values, though.

\subsection{Winds at very low metallicity}

As indicated in the introduction and as discussed in several contributions
throughout this symposium there is growing evidence that the
evolution of galaxies in the early universe is heavily influenced 
by the formation of first
generations of very massive stars. Thus, it is important to understand the
nature of radiation driven winds at very low metallicity. As has been shown
by \cite{kud02}, radiative line forces show a different dependence on
optical depth and electron density at very low metallicity, which requires
significant modfications of the theoretical description. With those
implemented it can be shown that the power law dependence on metallicity
of mass-loss rates, wind momenta and velocities breaks down and a much
stronger dependence on metallicity is found. For details we refer the reader 
to the paper by \cite{kud02}, which also provides ionizing fluxes and
predicted UV spectra as a function of metallicity. 

\subsection{The effects of rotation and instabilities}

There are many mechanisms, which might lead to an enhancement of mass-loss
during stellar evolution. A classical mechanism already taken into account by
\cite{pauldrach86} and \cite{friend86} is through the centrifugal forces provided
by stellar rotation. \cite{kupu00}, \cite{petrenz00}, and \cite{owocki05}
give an overview about the possible effects. In particular at low
metallicities, when massive O-stars are hotter and have much smaller radii
and the effects of radiation driven winds become smaller, stellar rotation
might become a crucial mechanism not only for stellar mass-loss, but also
for stellar evolution (see \cite{marigo03}, \cite{maeder05},
\cite{meynet05}, \cite{hirschi05}, \cite{ciappini06}, \cite{meynet06}).

Continuum driven winds and their instabilities as very likely encountered in
LBV outbursts are another important physically mechanism at least for
objects very close to the Eddington limit (see \cite{shaviv01}, \cite{shaviv05},
\cite{owocki04}, \cite{owocki05}, \cite{smith06} and the contribution by
Nathan Smith at these proceedings).

Stellar pulsations have also been discussed frequently to enhance stellar
mass-loss, but as shown by \cite{baraffe01} they become less important at
low metallicities.

\subsection{The problem of the ``weak wind stars''}

UV and optical studies of O-stars in the SMC with luminosities $L \le 10^{5.5}
L_{\odot}$ by \cite{bouret03} and \cite{martins04} indicated much
smaller stellar wind momenta than expected from the theory of radiation
driven winds and from a simple extrapolation of the WLR from higher
luminosities. More recently, \cite{martins05b} have investigated similar ``weak
wind stars'' in the Milky Way and it was found that O-dwarfs with $L \le
10^{5.2} L_{\odot}$ indeed seem to have much weaker winds than predicted by
the theory with a discrepancy of the order of a factor of hundred. It is
always simple to speculate about too strong stellar winds. One can invent
additional wind driving mechanisms or blame the neglect of clumping in
the diagnostics as the reason. However, it is very difficult to explain
winds that are too weak, because the radiative force at a given luminosity
is always there and cannot be simply switched off. Thus, it is natural to be
suspicious about the accuracy of the spectroscopic diagnostics of these
objects.

The determination of mass-loss rates comes mostly from the analysis of UV
metal lines, since H$_{\alpha}$ provides only upper limits in most of the
cases. While the diagnostics of those lines have been done in the most
careful way with the state-of-the-art model atmospheres described before,
the mass-loss rates determined depend crucially on ionization calculations
which could severely be affected by the soft X-ray emission of shocks in the
stellar wind flow. The authors are aware of this problem and have included
effects of shock emission, but it is open at this point whether this
treatment of shocks is sufficient. Other possible options - from our point
of view less likely - to explain the weak wind stars are discussed by 
\cite{martins05b} and \cite{mokiem06b}.

\subsection{Stellar wind clumping}

\begin{figure}
\centerline{\hbox{
   \psfig{figure=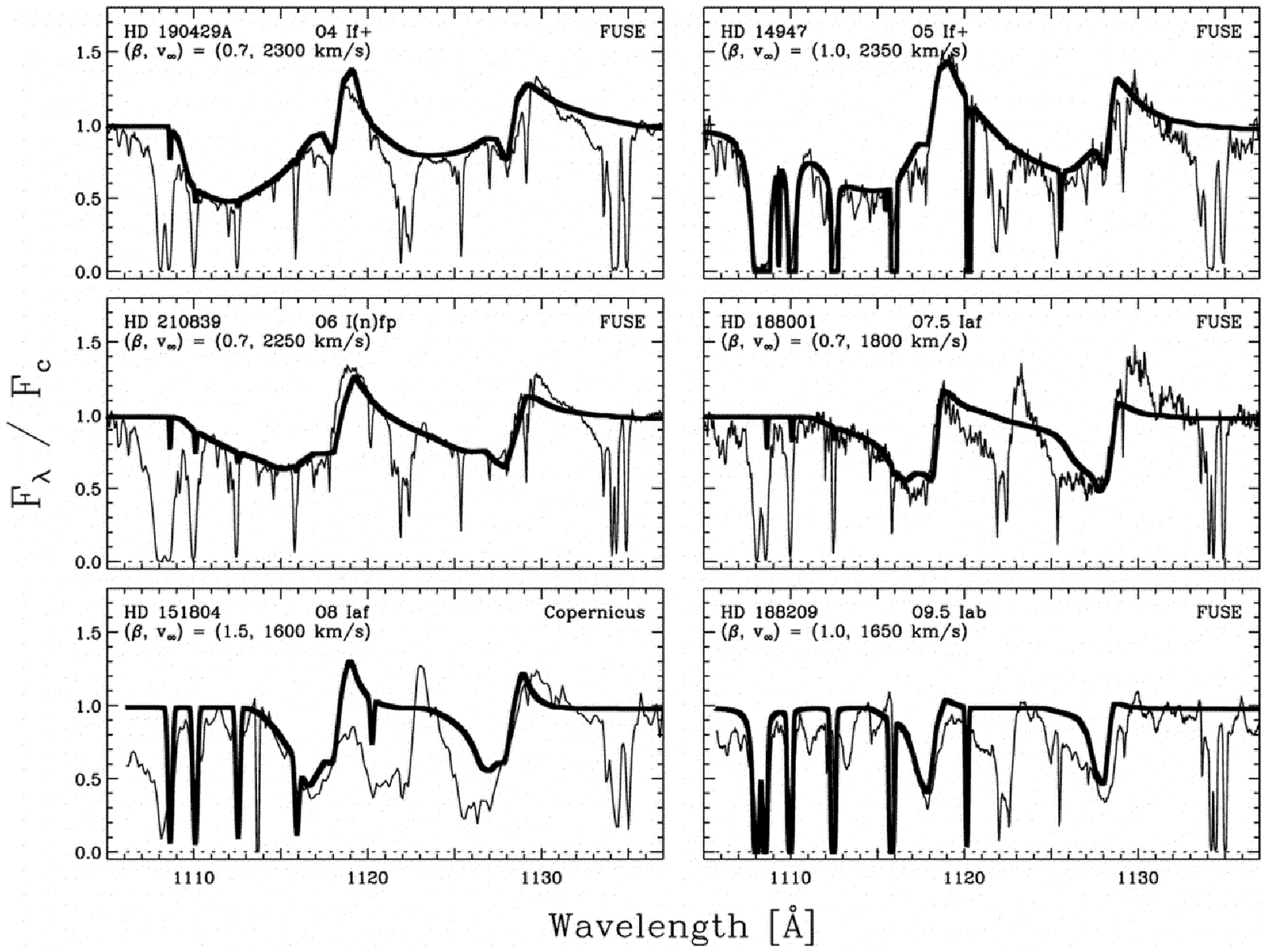,width=7cm}
   \psfig{figure=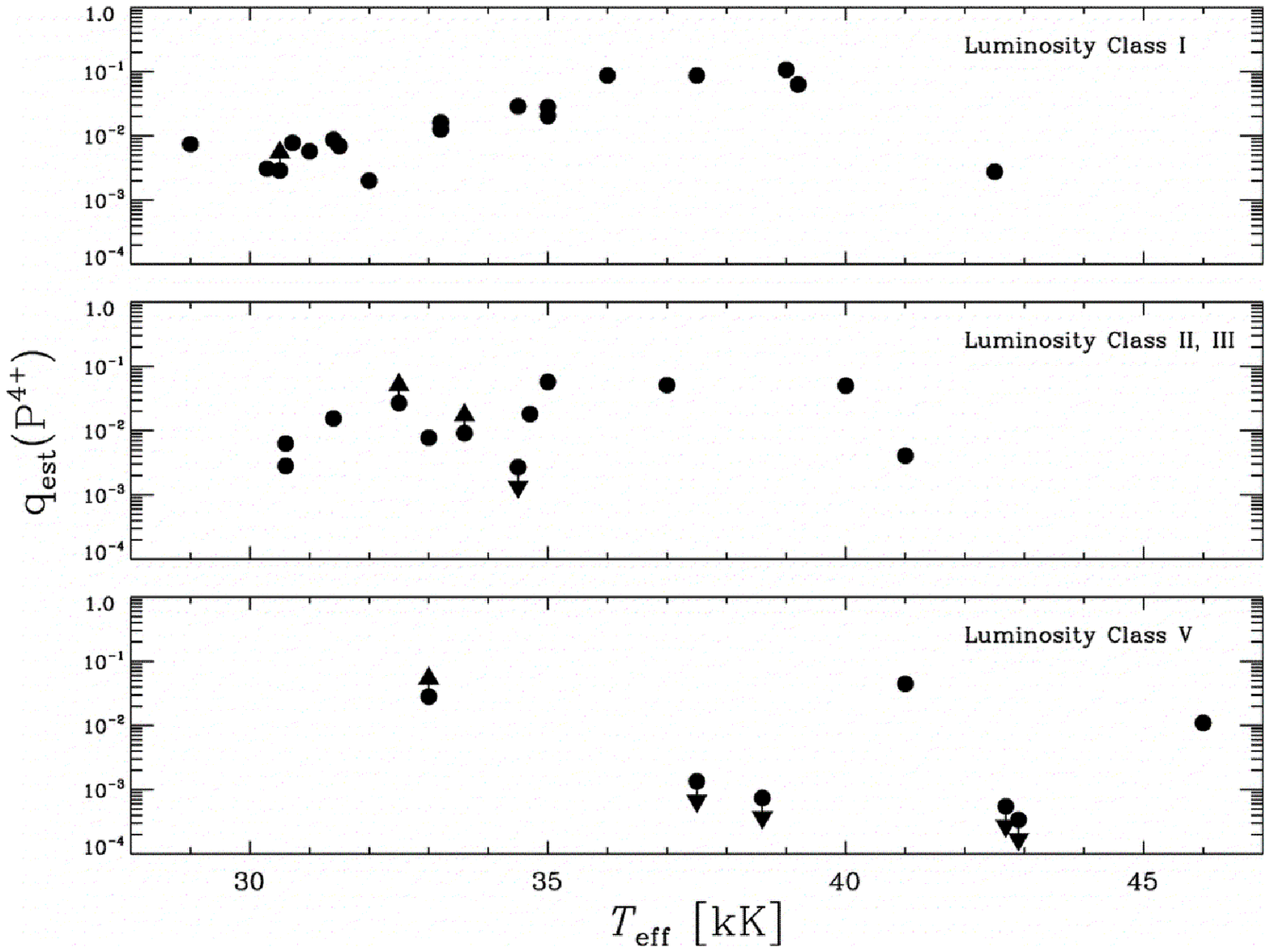,width=7cm}
         }}
\caption[]{
Left: Radiative transfer fits of the far UV PV lines of MW O-stars. 
Right: Ionization fraction of PV as determined from fits of H$_{\alpha}$ and
PV as function of effective temperature. (From \cite{fullerton06}).  
\label{pv}}
\end{figure}

While H$_{\alpha}$ is, in principle, a perfect tool to measure mass-loss 
rates (see \cite{kupu00}, for discussion and 
references), the results might be affected by stellar wind clumping. It has
been known since long that line driven winds are intrinsically unstable
(Owocki et al. 1988, 2004). This might lead to inhomogeneous, clumped winds
such as described by \cite{owocki02} with regions of enhanced
density $\rho_{cl}$ and regions, where the density is much lower. In a very
simple description, introducing clumping factors $f_{cl}$ similar as in
PN diagnostics, the relationship between the average density of the stellar
wind flow $\rho_{av}$ and the density in the clumps is then given by
$\rho_{cl} = \rho_{av} f_{cl}$. The  same relationship holds for the
occupation numbers $n_{i}$ of ions.

Line opacities $\kappa$ depend on density through $\kappa \propto n_{i}
\propto \rho^{x}$ and for very small, optically thin clumps the avarage 
optical line depth in the wind is given by 
$\tau_{av} \propto n^{av}_{i} \propto n^{cl}_{i} f^{-1} \propto
\rho_{av}^{x} f^{x-1}$. For a dominating ionization stage we have $x = 1$ 
and the clumping along the line of sight cancels and does not affect the 
diagnostics. However, bound hydrogen is a minor ionization stage in hot 
stars depending on recombination from ionized hydrogen 
with $n_{i}(H) \propto n_{E}n_{P} \propto \rho^{2}$. Thus, 
if $f_{cl}$ is significantly larger than one, the H$_{\alpha}$
mass-loss rate diagnostic is systematically affected and we have
$\dot{M}(H_{\alpha}) = \dot{M}(true) f_{cl}^{1/2}$, following from the fact
that $\dot{M}(true) \propto \rho_{av}.$

\begin{figure}
\centerline{\hbox{
   \psfig{figure=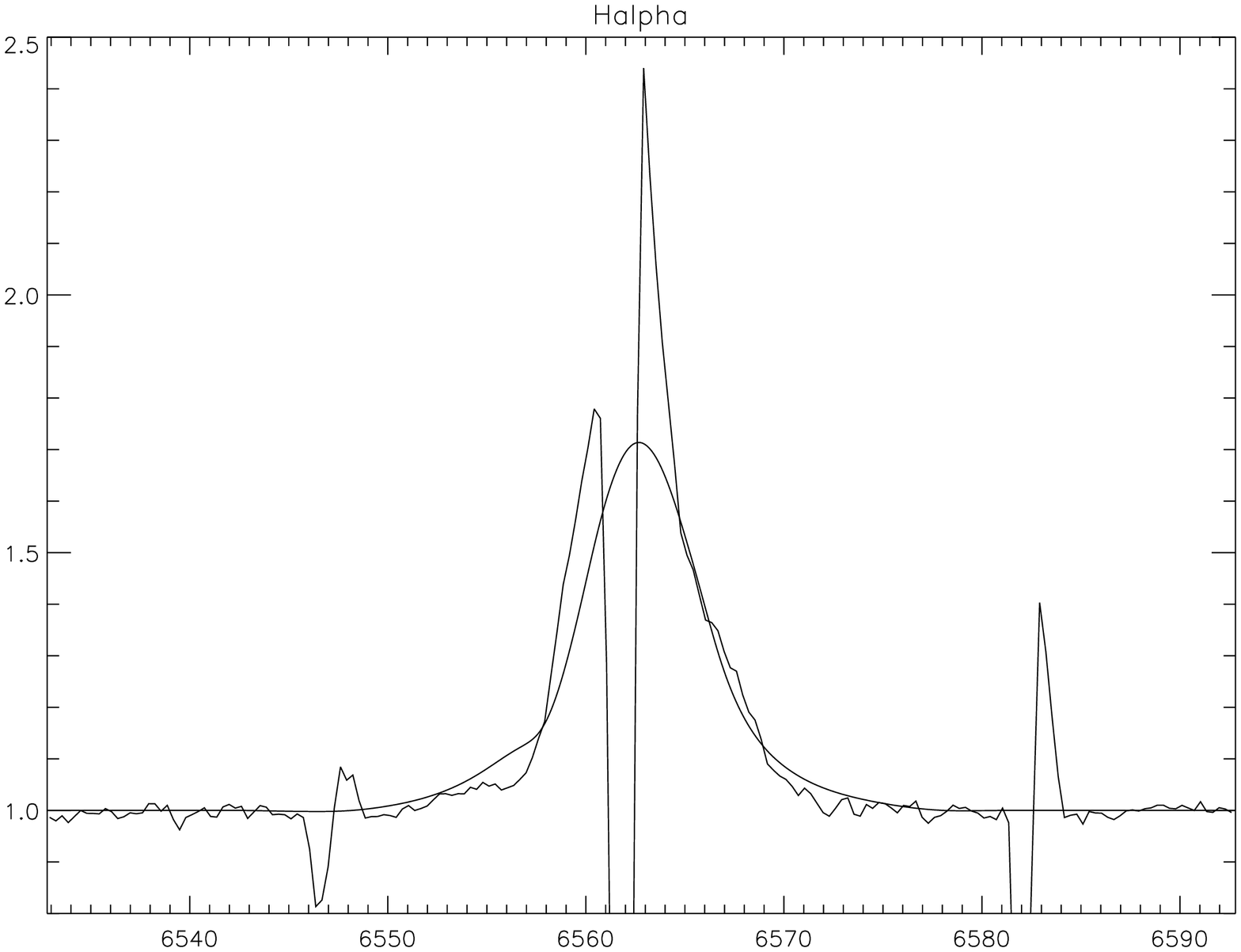,width=5cm,angle=90}
   \psfig{figure=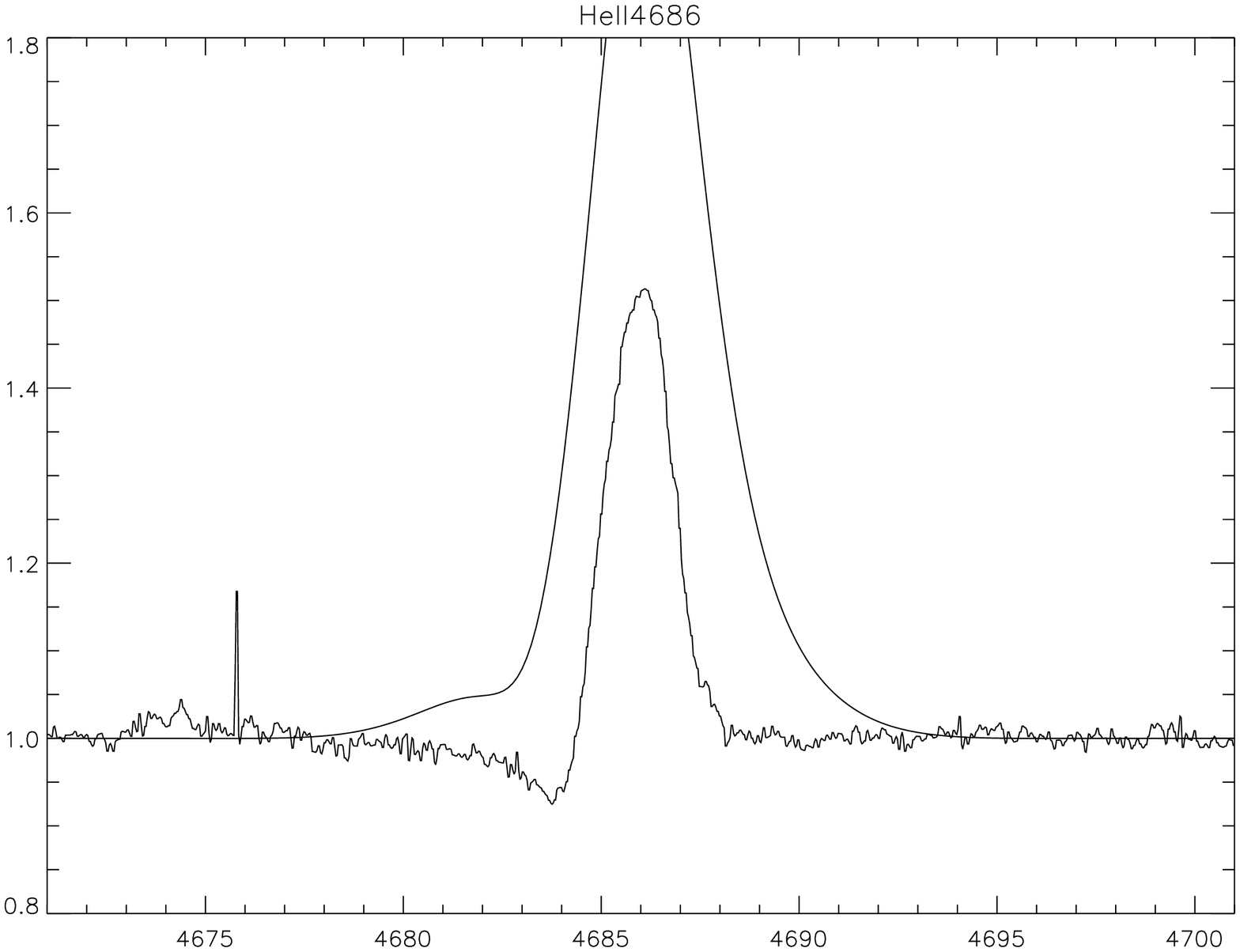,width=5cm,angle=90}
         }}
\centerline{\hbox{	 
   \psfig{figure=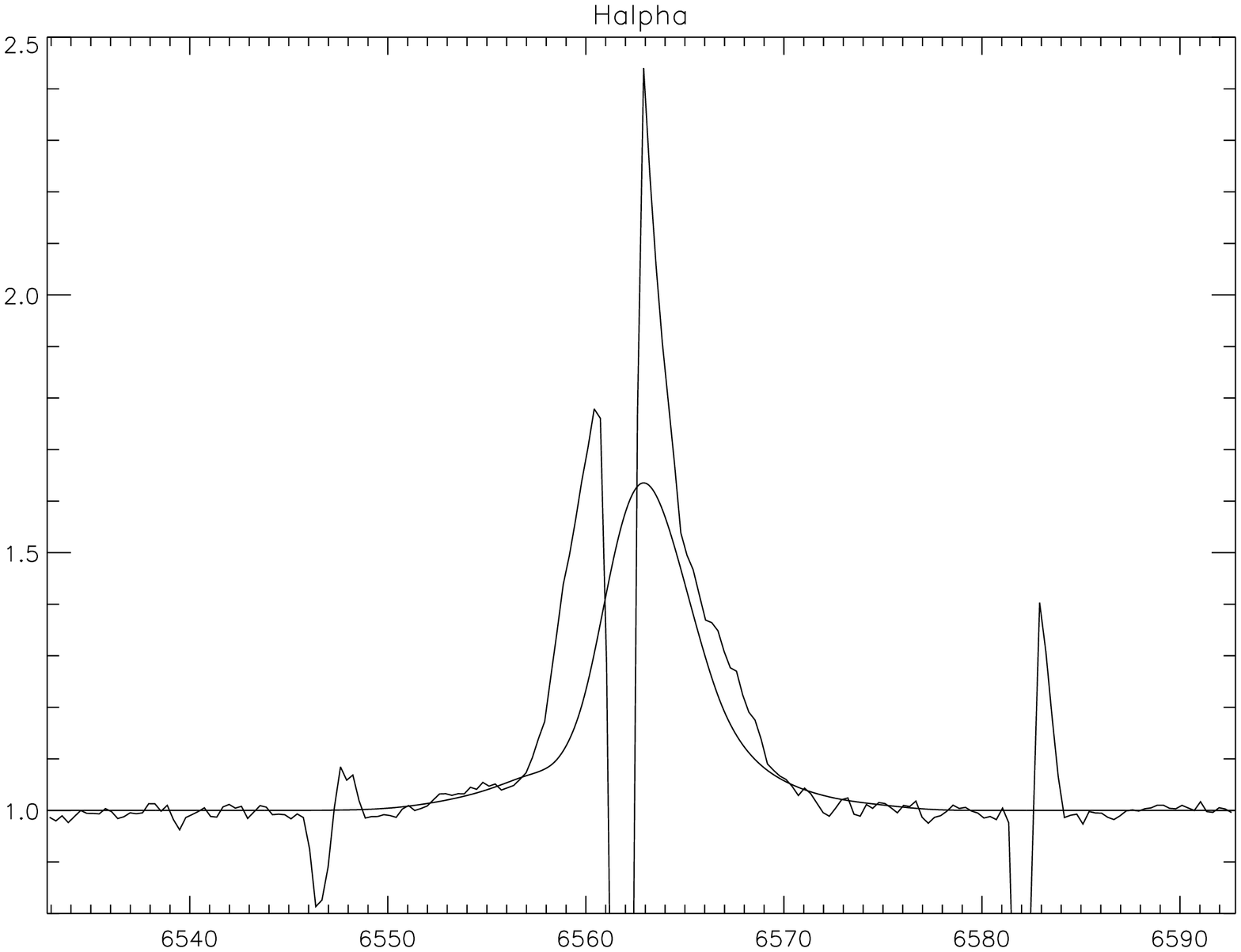,width=5cm,angle=90}
   \psfig{figure=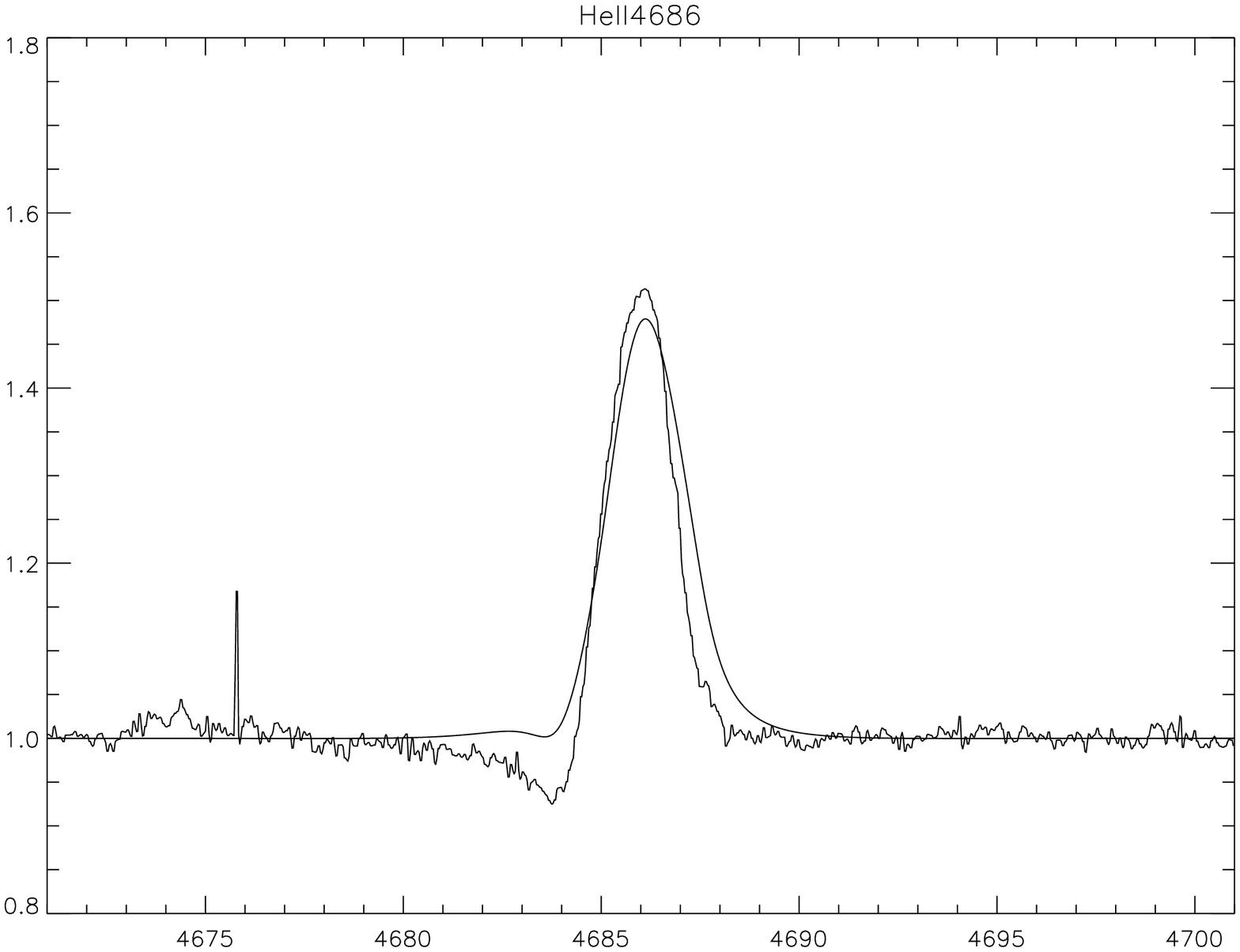,width=5cm,angle=90}
         }}
\centerline{\hbox{	 
   \psfig{figure=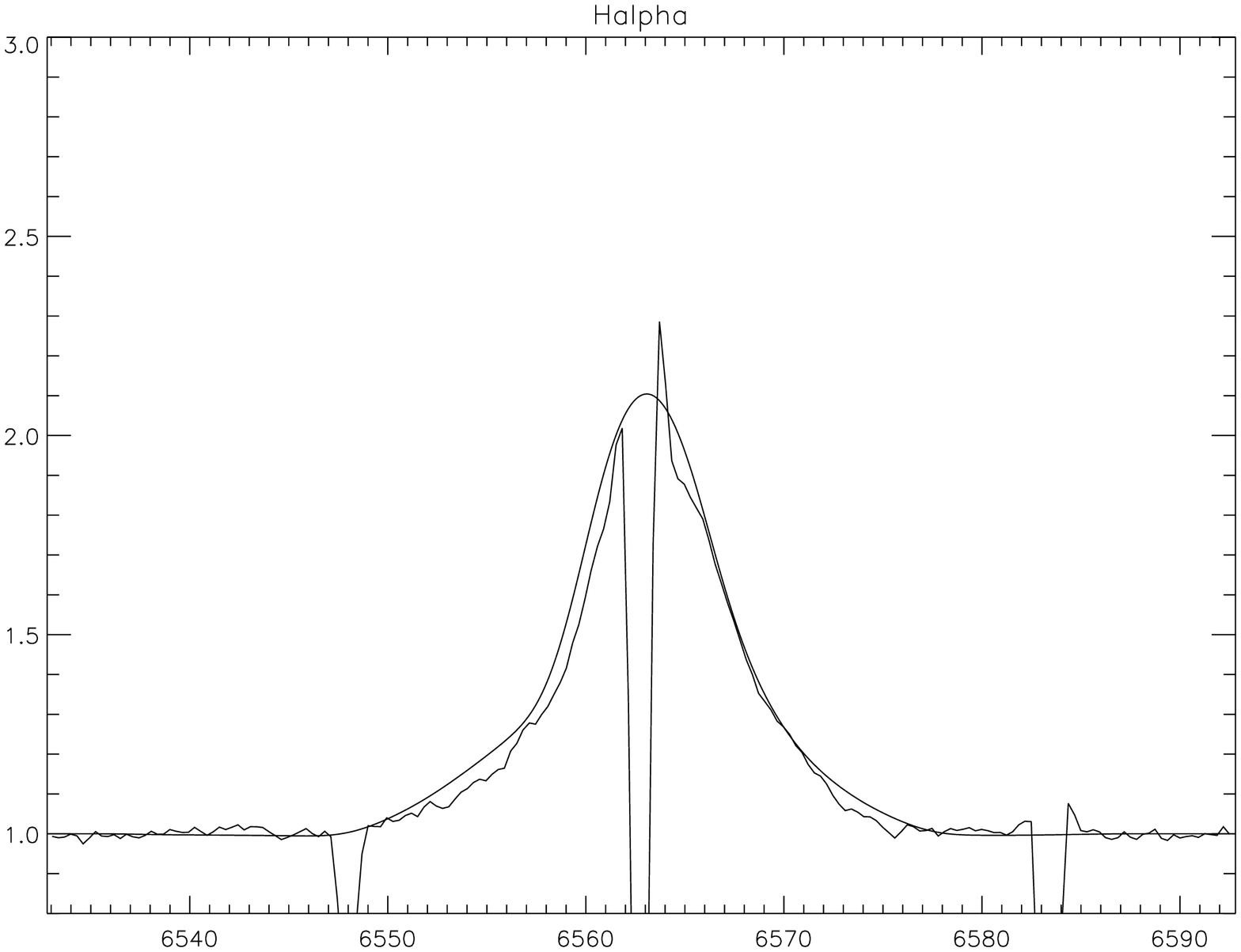,width=5cm,angle=90}
   \psfig{figure=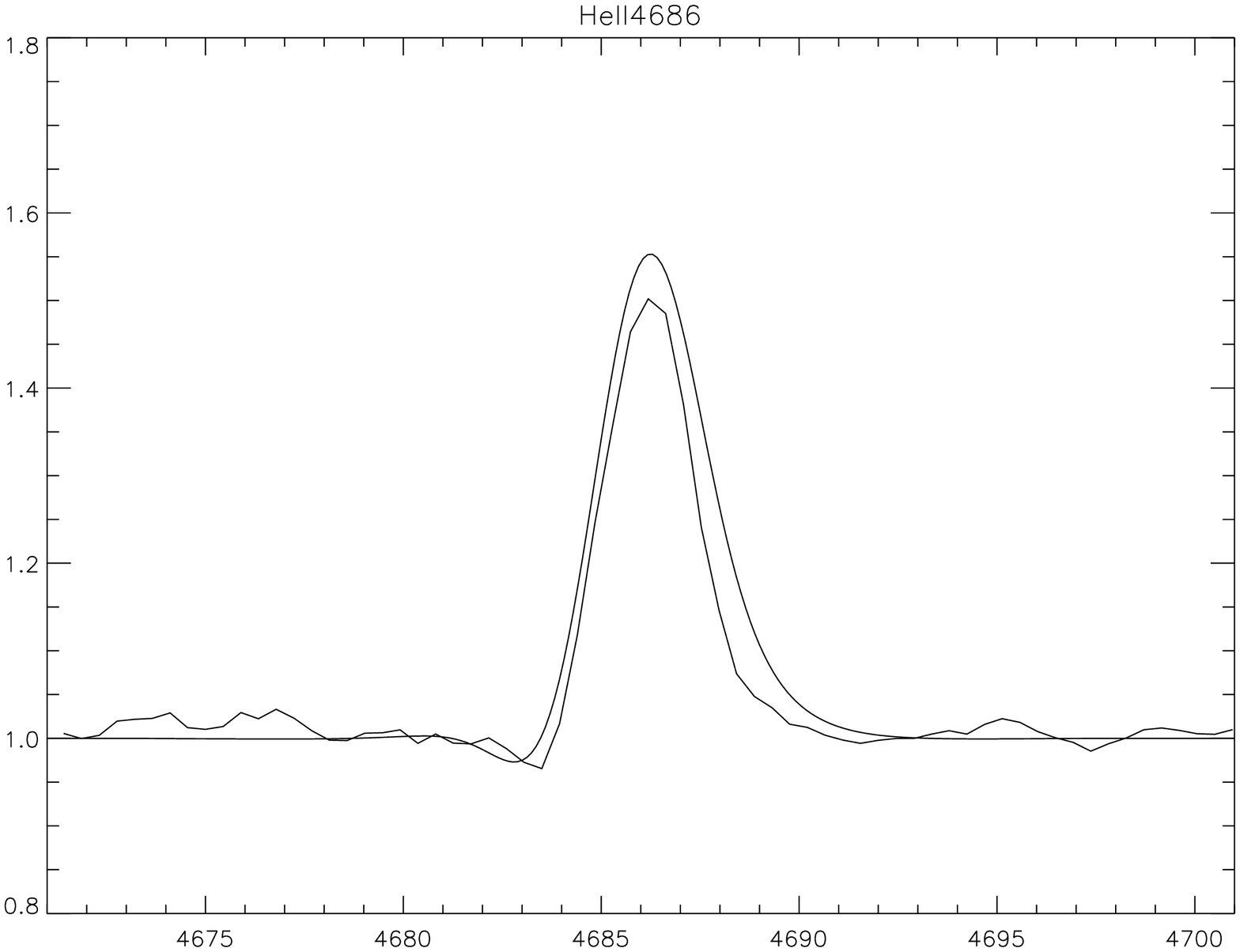,width=5cm,angle=90}
         }}	 
\caption[]{
Diagnostic of clumping factors in the winds of CSPN. The first two rows show
model fits of the H$_{\alpha}$ (left) and HeII 4686 (right) profile of IC418
with $f_{cl} = 1$ (top row) and $f_{cl} = 50$ indicating enormous clumping.
The bottom row shows the same for He2-108 and $f_{cl} = 1$ demonstrating
that for this object the wind is very likely homogeneous. Note that the
H$_{\alpha}$ and NII nebular emission lines have been (imperfectly) subtracted.
(From \cite{kud06a}). 
\label{heii}}
\end{figure}

The spectral diagnostics of clumping is difficult. In principle, 
it requires the comparison of lines with different exponents x in the 
density dependence of their opacities. In WR-type stars with very dense winds 
and very strong wind emission lines incoherent electron scattering produces 
wide emission wings, the strength of which goes with $x \sim 1$. Clumping 
factors of the order of ten to twenty were found (\cite{hillier91}, see also
contribution by Paul Crowther in these proceedings). 
This technique does not work for O-type stars, as their winds
have much lower density. Also the UV P-Cygni lines of dominating ions
provide usually little help, as these lines are mostly saturated and the
ionization equlibria are uncertain. However, in most recent work on massive
O-stars using FUSE and Copernicus spectra the PV resonance line at 1118 and
1128 $\AA$ has been used as an indicator of clumping. The advantage of PV is
the low cosmic abundance so that the line is completely unsaturated even when
in a dominating ionization stage. Substantial clumping was found 
(\cite{hillier03}, Bouret et al., 2003, 2005) with clumping factors of the
order of ten.

Very recently, \cite{fullerton06} have carried out a comprehensive study of
Milky Way O-stars with well observed FUSE PV line profiles producing
detailed PV radiative transfer line fits to determine the
product $q(P^{4+})\dot{M}$, where $q(P^{4+})$ is the ionization fraction of PV
in the ground-state (see Fig.\ref{pv}). Comparing with mass-loss rates derived
from H$_{\alpha}$ (or radio free-free emission) they produced the plot of
$q(P^{4+})$ as function of T$_{eff}$ also shown in Fig.\ref{pv}. Assuming
that PV is a dominating ionization stage at all temperatures (and, therefore,
$q(P^{4+})$ = 1), they conclude that the H$_{\alpha}$ mass-loss rates are
too high by a very large factor up to many orders of magnitudes and implying
enormous filling factors.

This is very important work based on the best available diagnostic
techniques and poses a very serious problem. The crucial assumption is, of
course, $q(P^{4+})$ = 1. Test calculations done by us with the model
atmosphere code FASTWIND show that models (without shock emission) predict
$q(P^{4+})$ = 1 only for T$_{eff} \le 35kK$, which would imply a much lower
effect then at hotter temperatures. Clearly, for future work a very detailed
investigation of the PV ionization (including shock emission) is needed to
address this fundamental problem of clumping in O-star winds. The results
found by \cite{fullerton06} are certainly alarming.

There is an alternative method for the diagnostics of clumping at least for
cool O-stars with T$_{eff} \le 37,000$K, where HeII is a dominant ionization
stage. That means for objects with strong winds and HeII 4686 in emission
and formed in the wind this line should have a density dependence close to 
$x =1 $. Its relative strenght to H$_{\alpha}$ should allow to constrain 
f$_{cl}$.

\cite{kud06a} have applied this technique to study clumping in the winds of 
CSPN with very interesting results (see Fig.~\ref{heii}) yielding clumping
factors varying in a range from 50 to 1.


\begin{thebibliography}{}

  \bibitem[Abbott (1982)]{abbott82}
  \textsc{Abbot, D.C.} 1982
  \textit{ApJ} \textbf{259}, 282

  \bibitem[Abbott \& Hummer (1985)]{abbott85}
  \textsc{Abbot, D.C. \& Hummer, D.G.} 1985
  \textit{ApJ} \textbf{294}, 286

  \bibitem[Baraffe et al. (2001)]{baraffe01}
  \textsc{Baraffe, I., Heger, A., Woosley, S.E.} 2001
  \textit{ApJ} \textbf{550}, 890

  \bibitem[Barton et al. (2004)]{barton04}
  \textsc{Barton, E., Dave, R., Smith, J. et al.} 2004
  \textit{ApJ} \textbf{604}, L1

  \bibitem[Bianchi \& Garcia (2002)]{bianchi02}
  \textsc{Bianchi, L. \& Garcia, M.} 2002
  \textit{ApJ} \textbf{581}, 610

  \bibitem[Bouret et al. (2003)]{bouret03}
  \textsc{Bouret, J.C., Lanz, T., Hillier, J.D. et al.} 2003
  \textit{ApJ} \textbf{595}, 1182
  
  \bibitem[Bouret et al. (2005)]{bouret05}
  \textsc{Bouret, J.C., Lanz, T., Hillier, J.D.} 2005
  \textit{A\&A} \textbf{438}, 301

  \bibitem[Bresolin (2003)]{bres03}
  \textsc{Bresolin, F.} 2003
  \textit{Lect. Notes Phys.} \textbf{635}, 149

  \bibitem[Bresolin et al. (2001)]{bres01}
  \textsc{Bresolin, F., Kudritzki, R.P., Mendez, R.H., Przybilla, N.} 2001
  \textit{ApJ} \textbf{548}, L149
  
  \bibitem[Bresolin et al. (2002a)]{bres02a}
  \textsc{Bresolin, F., Gieren, W., Kudritzki, R.P., Pietrzynski, G.,
  Przybilla, N.} 2002a
  \textit{ApJ} \textbf{567}, 227

  \bibitem[Bresolin et al. (2002b)]{bres02b}
  \textsc{Bresolin, F., Kudritzki, R.P., Najarro, F., Gieren, W., 
  Pietrzynski, G.} 2002b
  \textit{ApJ} \textbf{577}, L107
  
  \bibitem[Bresolin et al. (2004)]{bres04}
  \textsc{Bresolin, Pietrzynski, G., Gieren, W., Kudritzki, R.P., 
  Przybilla, N., Fouque, P.} 2004
  \textit{ApJ} \textbf{600}, 182
  
  \bibitem[Bresolin et al. (2006)]{bres06}
  \textsc{Bresolin, Pietrzynski, G., Urbaneja, M.A., Gieren, W., 
  Kudritzki, R.P., Venn, K.A.} 2006
  \textit{ApJ} \textbf{600}, 182

  \bibitem[Bromm, Kudritzki \& Loeb (2001)]{bromm01}
  \textsc{Bromm, V., Kudritzki, R.P., Loeb, A.} 2001
  \textit{ApJ}, in press

  \bibitem[Ciappini et al. (2006)]{ciappini06}
  \textsc{Ciappini, C., Hirschi, R., Meynet, G. et al.} 2006
  \textit{A\&A} \textbf{449}, L27

  \bibitem[Crowther et al. (2002)]{crowther02}
  \textsc{Crowther, P.A., Hillier, D.J., Evans, C.J., Fullerton, A.W., de
Marco, O., Willis, A.J.} 2004
  \textit{ApJ} \textbf{579}, 774

  \bibitem[Crowther, Lennon \& Walborn(2005)]{crowther05}
  \textsc{Crowther, P.A., Lennon, D.J., Walborn, N.R.} 2005
  \textit{A\&A} \textbf{446}, 279

  \bibitem[Evans \& Howarth (2003)]{evans03}
  \textsc{Evans, C.J., Howarth, I.} 2003
  \textit{MNRAS} \textbf{345}, 1223

  \bibitem[Evans et al. (2004)]{evans04}
  \textsc{Evans, C.J., Crowther, P.A., Fullerton, A.W., \& Hillier, D.J.} 2004
  \textit{ApJ} \textbf{610}, 1021

  \bibitem[Figer et al. (1998)]{figer98}
  \textsc{Figer, D.F., Najarro, F., Morris, M. et al.} 1998
  \textit{ApJ} \textbf{506}, 384

  \bibitem[Friend \& Abbott (1986)]{friend86}
  \textsc{Friend, D. \& Abbott, D.C.} 1986
  \textit{ApJ} \textbf{202}, 153

  \bibitem[Fullerton et al. (2006)]{fullerton06}
  \textsc{Fullerton, A., Massa, D., Prinja, R.} 2006
  \textit{ApJ}, in press

  \bibitem[Gabler, Gabler, Kudritzki et al. (1989)]{gab89}
  \textsc{Gabler, R., Gabler, A., Kudritzki, R.P., Puls, J., Pauldrach, A.W.A.} 1989
  \textit{A\&A} \textbf{226}, 162

  \bibitem[Gabler, Kudritzki \& Mendez (1991)]{gab91}
  \textsc{Gabler, R., Kudritzki, R.P., Mendez, R.H.} 1991
  \textit{A\&A} \textbf{245}, 587
  
  \bibitem[Gabler, Gabler, Kudritzki et al. (1992)]{gab92}
  \textsc{Gabler, R., Gabler, A., Kudritzki, R.P., Mendez, R.H.} 1992
  \textit{A\&A} \textbf{265}, 656

  \bibitem[Garcia \& Bianchi (2004)]{garcia04}
  \textsc{Garcia, M. \& Bianchi, L.} 2004
  \textit{ApJ} \textbf{606}, 497

  \bibitem[Hanson, Kudritzki, Kenworthy et al. (2005)]{hanson05}
  \textsc{Hanson, M.M., Kudritzki, R.P., Kenworthy, M.A., Puls, J.,
Tokunaga, A.T.} 2005
  \textit{ApJS} \textbf{161}, 154

  \bibitem[Heap, Lanz \& Hubeny (2006)]{heap06}
  \textsc{Heap, S.R., Lanz, T., Hubeny, I.} 2006
  \textit{ApJ} \textbf{638}, 409

  \bibitem[Herrero, Kudritzki, et al. (1992)]{herrero92}
     \textsc{Herrero, A., Kudritzki, R.P., Vilchez, J.M. et al.} 1992
     \textit{A\&A} \textbf{261}, 209 

  \bibitem[Herrero et al. (2002)]{herrero02}
     \textsc{Herrero, A., Puls, J., Najarro, F.} 2002
     \textit{A\&A} \textbf{396}, 949

  \bibitem[Hillier (1991)]{hillier91}
  \textsc{Hillier, J.D.} 1991
  \textit{A\&A} \textbf{247}, 455

  \bibitem[Hillier et al. (2003)]{hillier03}
  \textsc{Hillier, J.D., Lanz, T., Heap, S.R. et al.} 2003
  \textit{ApJ} \textbf{588}, 1039

  \bibitem[Hirschi et al. (2005)]{hirschi05}
  \textsc{Hirschi, R., Meynet, G., Maeder, A.} 2005
  \textit{A\&A} \textbf{443}, 581

  \bibitem[Hummer (1982)]{hummer82}
  \textsc{Hummer, D.G.} 1982
  \textit{ApJ} \textbf{257}, 724

  \bibitem[Kaufer et al. (2004)]{kaufer04}
     \textsc{Kaufer, A., Venn, K.A., Tolstoy, E., Pinte, C.,
     Kudritzki, R.P.} 2004
     \textit{AJ} \textbf{127}, 2723

  \bibitem[Kudritzki (1980)]{kud80}
     \textsc{Kudritzki, R.P.} 1980
     \textit{A\&A} \textbf{85}, 174

  \bibitem[Kudritzki (1998)]{kud98}
     \textsc{Kudritzki, R.P.} 1998
     {Proc. ``Stellar Astrophysics for the Local Group'', eds. A. Aparicio,
     A. Herrero \& F. Sanchez}
     \textit{VIII Canary Islands Winter School in Astrophysics, Cambridge
     University Press}, 149--261

  \bibitem[Kudritzki (2000)]{kud00}
     \textsc{Kudritzki, R.P.} 2000
     \textit{STScI Symposium Series} \textbf{Vol. 12}, 202

   \bibitem[Kudritzki (2002)]{kud02}
     \textsc{Kudritzki, R.P.} 2002
     \textit{ApJ} \textbf{577}, 389

  \bibitem[Kudritzki et al. (1983)]{kud83}
     \textsc{Kudritzki, R.P., Simon, K.P., \& Hamann, W.R.} 1983
     \textit{A\&A} \textbf{118}, 245

  \bibitem[Kudritzki et al. (1987)]{kud87}
     \textsc{Kudritzki, R.P., Pauldrach, A.W.A., Puls, J.} 1987
     \textit{A\&A} \textbf{173}, 293

  \bibitem[Kudritzki et al. (1989)]{kud89}
     \textsc{Kudritzki, R.P., Cabanne, K.P., Husfeld, D. et al.} 1989
     \textit{A\&A} \textbf{226}, 235   

  \bibitem[Kudritzki \& Hummer (1990)]{kuhu90}
     \textsc{Kudritzki, R.P., Hummer, D.G.} 1990
     \textit{AARA} \textbf{28}, 303

  \bibitem[Kudritzki, Lennon \& Puls (1995)]{kud95}
     \textsc{Kudritzki, R.P., Lennon, D.J., Puls, J.} 1995
     \textit{Science with the VLT, eds. J.R. Walsh, J. Danziger}, 246   

  \bibitem[Kudritzki et al. (1999)]{kud99}
     \textsc{Kudritzki, R.P., Puls, J, Lennon, D.J. et al.} 1999
     \textit{A\&A} \textbf{350}, 970   

  \bibitem[Kudritzki \& Puls (2000)]{kupu00}
     \textsc{Kudritzki, R.P., Puls, J.} 2000
     \textit{AARA} \textbf{38}, 613--666

  \bibitem[Kudritzki et al. (2003a)]{kud03a}
     \textsc{Kudritzki, R.P. and the GSMT Science Working Group} 2003a
     \textit{GSMT Science Working Group Report},
     http://www.aura-nio.noao.edu/

  \bibitem[Kudritzki (2003)]{kud03b}
  \textsc{Kudritzki, R.P.} 2003
  \textit{Lect. Notes Phys.} \textbf{635}, 123
  
  \bibitem[Kudritzki, Bresolin, \& Przybilla (2003)]{kud03c}
  \textsc{Kudritzki, R.P., Bresolin, F., Przybilla, N.} 2003b
  \textit{ApJ} \textbf{582}, L83

  \bibitem[Kudritzki, Urbaneja, \& Puls (2006a)]{kud06a}
  \textsc{Kudritzki, R.P., Urbaneja, M.A., \& Puls, J.} 2006a
  \textit{Proc. IAU Symposium No. 234, eds. M.J. Barlow \& R.H. Mendez}, 
  invited paper, in press

  \bibitem[Kudritzki, Urbaneja, Bresolin et al. (2006b)]{kud06b}
  \textsc{Kudritzki, R.P., Urbaneja, M.A., Bresolin, et al.} 2006b
  \textit{ApJ}, in prep.

  \bibitem[Lanz \& Hubeny (2003)]{lanz03}
  \textsc{Lanz, T., Hubeny, I.} 2003
  \textit{ApJ} \textbf{465}, 359

  \bibitem[Leitherer et al. (1992)]{leitherer92}
  \textsc{Leitherer, C., Robert., C. \& Drissen, L.} 1992
  \textit{ApJ} \textbf{401}, 596

  \bibitem[Lennon, D.J. (1997)]{lennon97}
     \textsc{Lennon} 1997
     \textit{A\&A} \textbf{317}, 871

  \bibitem[Lenorzer et al. (2004)]{lenorzer04}
     \textsc{Lenorzer, A., Mokiem, M.R., deKoter, A., Puls, J.} 2004
     \textit{A\&A} \textbf{422}, 275 

 \bibitem[Maeder et al. (2005)]{maeder05}
     \textsc{Maeder, A., Meynet, G., \& Hirschi, R.} 2005
     \textit{ASP Conf. Series} \textbf{332}, 3

\bibitem[Marigo et al. (2003)]{marigo03}
     \textsc{Marigo, P., Chiosi, C., Kudritzki, R.P.} 2003
     \textit{A\&A} \textbf{399}, 617

  \bibitem[Markova et al. (2004)]{markova04}
     \textsc{Markova, N., Puls, J., Repolust, T., Markov, H.} 2004
     \textit{A\&A} \textbf{413}, 693

  \bibitem[Martins et al. (2002)]{martins02}
  \textsc{Martins, F., Schaerer, D., Hillier, D.J.} 2002
  \textit{A\&A} \textbf{382}, 999

  \bibitem[Martins et al. (2004)]{martins04}
  \textsc{Martins, F., Schaerer, D., Hillier, D.J., \& Heydari-Malayeri, M.}
2004
  \textit{A\&A} \textbf{420}, 1087
  
  \bibitem[Martins et al. (2005a)]{martins05a}
  \textsc{Martins, F., Schaerer, D., Hillier, D.J.} 2005a
  \textit{A\&A} \textbf{436}, 1049
  
  \bibitem[Martins et al. (2005b)]{martins05b}
  \textsc{Martins, F., Schaerer, D., Hillier, D.J. et al.} 2005b
  \textit{A\&A} \textbf{441}, 735

  \bibitem[Massey, Bresolin, Kudritzki et al. (2004)]{massey04}
  \textsc{Massey, P., Bresolin, F., Kudritzki, R.P., Puls, J. \& Pauldrach,
A.W.A.} 2004
  \textit{ApJ} \textbf{608}, 1001
  
  \bibitem[Massey, Puls, Pauldrach et al. (2005)]{massey05}
  \textsc{Massey, P., Puls, J., Pauldrach, A.W.A., Bresolin, F., Kudritzki, R.P., 
  \& Simon, T.} 2005
  \textit{ApJ} \textbf{627}, 477

  \bibitem[McErlean et al. (1998)]{mcerlean98}
     \textsc{McErlean, N.D., Lennon, D.J., Dufton, P.L.} 1998
     \textit{A\&A} \textbf{329}, 613

  \bibitem[McErlean et al. (1999)]{mcerlean99}
     \textsc{McErlean, N.D., Lennon, D.J., Dufton, P.L.} 1999
     \textit{A\&A} \textbf{349}, 553

  \bibitem[Meynet et al. (1994)]{meynet94}
     \textsc{Meynet, G., Maeder, A., Schaller, G., Schaerer, D., 
     Charbonnel, C.} 1994
     \textit{A\&AS} \textbf{103}, 97

  \bibitem[Meynet \& Maeder (2000)]{meynet00}
     \textsc{Meynet, G. \& Maeder, A.} 2000
     \textit{A\&A} \textbf{361}, 101

  \bibitem[Meynet et al. (2005)]{meynet05}
     \textsc{Meynet, G., Maeder, A., \& Ekstr\"om, S.} 2005
     \textit{ASP Conf. Series} \textbf{332}, 228

  \bibitem[Meynet et al. (2006)]{meynet06}
     \textsc{Meynet, G., Ekstr\"om, S., \& Maeder, A.} 2006
     \textit{A\&A} \textbf{447}, 623

 \bibitem[Mokiem et al. (2004)]{mokiem04}
     \textsc{Mokiem, M.R., Martin-Hernandez, N.L., Lenorzer, A., deKoter,
A., Tielens, A.G.G.M.} 2004
     \textit{A\&A} \textbf{419}, 319

  \bibitem[Mokiem et al. (2005)]{mokiem05}
     \textsc{Mokiem, M.R., deKoter, A., Puls, J. et al.} 2005
     \textit{A\&A} \textbf{441}, 711

  \bibitem[Mokiem et al. (2006a)]{mokiem06a}
     \textsc{Mokiem, M.R., deKoter, A., Evans, C. et al.} 2006a
     \textit{A\&A}, in press

  \bibitem[Mokiem et al. (2006b)]{mokiem06b}
     \textsc{Mokiem, M.R., deKoter, A., Evans, C.et al.} 2006b
     \textit{A\&A}, in press

  \bibitem[Najarro et al. (1994)]{naj94}
  \textsc{Najarro, F., Hillier, D.J., Kudritzki, R.P. et al.} 1994
  \textit{A\&A} \textbf{285}, 573

  \bibitem[Najarro, Kudritzki, Cassinelli et al. (1996)]{naj96}
  \textsc{Najarro, F., Kudritzki, Cassinelli, J.P., Stahl, O., Hillier, D.J.} 1996
  \textit{A\&A} \textbf{306}, 892
  
  \bibitem[Najarro et al. (1997)]{naj97}
  \textsc{Najarro, F., Krabbe, A., Genzel, R., Lutz, D., Kudritzki, R.P., 
  \& Hiller, D.J.} 1997
  \textit{A\&A} \textbf{325}, 700
  
  \bibitem[Najarro et al. (2004)]{naj04}
  \textsc{Najarro, F., Figer, D.F., Hillier, D.J., Kudritzki, R.P.} 2004
  \textit{ApJ} \textbf{611}, L105

   \bibitem[Owocki et al. 1988]{owocki88}
   \textsc{Owocki, S.P., Castor, J.I., Rybicki, G.B.} 1988 
   \textit{ApJ} \textbf{335}, 914

  \bibitem[Owocki \& Runacres (2002)]{owocki02}
  \textsc{Owocki, S.P., Runacres, M.C.} 2002 
  \textit{A\&A} \textbf{381}, 1015

  \bibitem[Owocki et al. (2004)]{owocki04}
     \textsc{Owocki, S., Gayley, K., Shaviv, N.} 2004
     \textit{ApJ} \textbf{616}, 520

  \bibitem[Owocki (2005)]{owocki05}
     \textsc{Owocki, S.} 2005
     \textit{ASP Conf. Series} \textbf{332}, 169

  \bibitem[Pauldrach, Puls, \& Kudritzki]{pauldrach86}
     \textsc{Pauldrach, A.W.A., Puls, J., Kudritzki, R.P.} 1986
     \textit{A\&A} \textbf{164}, 86

  \bibitem[Pauldrach, Lennon \& Hoffmann (2001)]{pauldrach01}
     \textsc{Pauldrach, A.W.A., Lennon, M., Hoffmann, T.} 2001
     \textit{A\&A} \textbf{375}, 161

  \bibitem[Petrenz \& Puls(2000)]{petrenz00}
     \textsc{Petrenz, P. \& Puls, J.} 2000
     \textit{A\&A} \textbf{358}, 956

  \bibitem[Prinja \& Massa (1998)]{prinja98}
     \textsc{Prinja, R.K., \& Massa, D.} 1998
     \textit{ASP Conf. Series} \textbf{131}, 218

  \bibitem[Przybilla (2002)]{przy02}
     \textsc{Przybilla, N.} 2002
     \textit{PhD Thesis, Ludwig-Maximilians-Universtit\"at M\"unchen}

  \bibitem[Przybilla et al. (2001a)]{przy01a}
     \textsc{Przybilla, N., Butler, K., Becker, S.R., Kudritzki, R.P.} 2001a
     \textit{A\&A} \textbf{369}, 1009

  \bibitem[Przybilla et al. (2001b)]{przy01b}
     \textsc{Przybilla, N., Butler, K., Kudritzki, R.P.} 2001b
     \textit{A\&A} \textbf{379}, 936
     
  \bibitem[Przybilla et al. (2001c)]{przy01c}
     \textsc{Przybilla, \& N., Butler, K.} 2001c
     \textit{A\&A} \textbf{379}, 995
     
  \bibitem[Przybilla et al. (2006)]{przy06}
     \textsc{Przybilla, N., Butler, K., Becker, S.R., Kudritzki, R.P.} 2006
     \textit{A\&A} \textbf{445}, 1099   

 \bibitem[Puls, Kudritzki, Herrero et al. (1996)]{puls96}
     \textsc{Puls, J., Kudritzki, R.P., Herrero, A. et al.} 1996
     \textit{A\&A} \textbf{305}, 171
 
  \bibitem[Puls et al. (2005)]{puls05}
     \textsc{Puls, J., Urbaneja, M.A., Venero, R. et al.} 2005
     \textit{A\&A} \textbf{435}, 669

   \bibitem[Repolust et al. (2004)]{repo04}
     \textsc{Repolust, T., Puls, J., Herrero, A.} 2004
     \textit{A\&A} \textbf{415}, 349

   \bibitem[Repolust et al. (2005)]{repo05}
     \textsc{Repolust, T., Puls, J., Hanson, M.M., Kudritzki, R.P., \&
    Mokiem, M.R.} 2005
     \textit{A\&A} \textbf{440}, 261

  \bibitem[Rix, Pettini, Leitherer et al. (2004)]{rix04}
     \textsc{Rix, S., Pettini, M., Leitherer, C., Bresolin, F., Kudritzki,
           R.P., Steidel, C.} 2004
     \textit{ApJ} \textbf{615}, 98   

  \bibitem[Schaerer (2003)]{schaerer03}
     \textsc{Schaerer, D.} 2003
     \textit{A\&A} \textbf{397}, 527

  \bibitem[Sellmaier et al. (1993)]{sellmaier93}
     \textsc{Sellmaier, F., Puls, J., Kudritzki, R.P., Gabler, A., \& 
     Gabler, R.} 1993
     \textit{A\&A} \textbf{273}, 533

  \bibitem[Shaviv (2001)]{shaviv01}
     \textsc{Shaviv, N..} 2001
     \textit{ApJ} \textbf{594}, 1093

  \bibitem[Shaviv (2005)]{shaviv05}
     \textsc{Shaviv, N..} 2005
     \textit{ASP Conf. Series} \textbf{332}, 180

  \bibitem[Smith \& Owocki (2006)]{smith06}
     \textsc{Smith, N. \& Owocki, S.} 2006
     \textit{ApJ}, submitted

  \bibitem[Trundle et al. (2004)]{trundle04}
     \textsc{Trundle, C., Lennon, D.J., Puls, J., \& Dufton, P.L.} 2004
     \textit{A\&A} \textbf{417}, 217

  \bibitem[Trundle \& Lennon (2005)]{trundle05}
     \textsc{Trundle, C., Lennon, D.J.} 2005
     \textit{A\&A} \textbf{434}, 677

  \bibitem[Urbaneja (2004)]{urba04}
     \textsc{Urbaneja, M.A.} 2004
     \textit{Thesis}, University of La Laguna, Istituto Astrofisica de 
     Canarias, Spain   
     
  \bibitem[Urbaneja et al. (2003)]{urba03}
     \textsc{Urbaneja, M.A., Herrero, A., Bresolin, F., Kudritzki, R.P.,
     Gieren, W., Puls, J.} 2003
     \textit{ApJ} \textbf{584}, L73
     
  \bibitem[Urbaneja et al. (2005a)]{urba05a}
     \textsc{Urbaneja, M.A., Herrero, A., Bresolin, F., Kudritzki, R.P.,
     Gieren, W., Puls, J., Przybilla, N., Najarro, F., Pietrzynski, G.} 2005a
     \textit{ApJ} \textbf{622}, 862
     
  \bibitem[Urbaneja et al. (2005b)]{urba05b}
     \textsc{Urbaneja, M.A., Herrero, A., Kudritzki, R.P., Najarro, F.,
     Smartt, S.J., Puls, J., Lennon, D.J., Corral, L.J.} 2005b
     \textit{ApJ} \textbf{635}, 311   

  \bibitem[Urbaneja, Kudritzki, Bresolin et al. (2006)]{urba06}
  \textsc{Urbaneja, M.A., Kudritzki, R.P., Bresolin, et al.} 2006
  \textit{ApJ}, in prep.

  \bibitem[Vacca et al. (1996)]{vacca96}
     \textsc{Vacca, W.D., Garmany, C.D., Shull, J.M.} 1996
     \textit{ApJ} \textbf{460}, 914 

  \bibitem[Venn (1999)]{venn99}
     \textsc{Venn, K.A.} 1999
     \textit{ApJ} \textbf{518}, 405
     
  \bibitem[Venn et al. (2000)]{venn00}
     \textsc{Venn, K.A., McCarthy, J.K., Lennon, D., Przybilla, N., 
     Kudritzki, R.P., Lemke, M. } 2000
     \textit{ApJ} \textbf{541}, 610   
     
  \bibitem[Venn et al. (2001)]{venn01}
     \textsc{Venn, K.A., Lennon, D., Kaufer, A., McCarthy, J.K., Przybilla,
     N., Kudritzki, R.P., Lemke, M. } 2001
     \textit{ApJ} \textbf{547}, 765   
     
  \bibitem[Venn et al. (2003)]{venn03}
     \textsc{Venn, K.A., Tolstoy, E., Kaufer, A., Skillman, E.D., Clarkson,
     S.M., Smartt, S.J., Lennon, D.J., Kudritzki, R.P.} 2003
     \textit{ApJ} \textbf{547}, 765     

  \bibitem[Vink et al. (2000)]{vink00}
     \textsc{Vink, J.S., de Koter, A., Lamers, H.J.G.L.M.} 2000
     \textit{A\&A} \textbf{362}, 295
     
  \bibitem[Vink et al. (2001)]{vink01}
     \textsc{Vink, J.S., de Koter, A., Lamers, H.J.G.L.M.} 2001
     \textit{A\&A} \textbf{369}, 574   

\end{thebibliography}
\end{document}